\documentclass[traditabstract]{aa}
\usepackage{natbib}
\usepackage{lscape}
\usepackage{wasysym}
\usepackage{txfonts}
\usepackage{graphicx}
\usepackage{hyperref}
\bibpunct{(}{)}{;}{a}{}{,} 
\usepackage{longtable}

\begin{document}

\title{
	The HARPS search for southern extra-solar planets.
	\thanks{Based on observations made with the HARPS instrument on the ESO 3.6-m telescope at La Silla Observatory (Chile), under programme IDs 072.C-0488 and 183.C-0972.}
	\thanks{Radial velocities are only available in the online version (Table 4, 5, 6, and 7) and in electronic form at the CDS via anonymous ftp to cdsarc.u-strasbg.fr (130.79.128.5) or via http://cdsweb.u-strasbg.fr/cgi-bin/qcat?J/A+A/}
	}

\subtitle{XXX. Planetary systems around stars with solar-like magnetic cycles and short-term activity variation}

\author{X. Dumusque\inst{1,2}
\and C. Lovis\inst{1}
\and D. S\'egransan\inst{1}
\and M. Mayor\inst{1}
\and S. Udry\inst{1}
\and W. Benz\inst{3}
\and F. Bouchy\inst{4,5}
\and G. Lo Curto\inst{6}
\and C. Mordasini\inst{7,3}
\and F. Pepe\inst{1}
\and D. Queloz\inst{1}
\and N. C. Santos\inst{2,8}
\and D. Naef\inst{1}
}

\institute{Observatoire de Gen\`eve, Universit\'e de Gen\`eve, 51 ch. des Maillettes, CH-1290 Versoix, Switzerland \\
\email{xavier.dumusque@unige.ch}
\and Centro de Astrof{\'\i}sica, Universidade do Porto, Rua das Estrelas, 4150-762 Porto, Portugal
\and Physikalisches Institut, Universit\"at Bern, Sidlerstrasse 5, CH-3012 Bern, Switzerland
\and Institut d'Astrophysique de Paris, UMR7095 CNRS, Universit\'e Pierre \& Marie Curie, 98bis Bd Arago, F-75014 Paris, France
\and Observatoire de Haute-Provence, CNRS/OAMP, F-04870 St. Michel l'Observatoire, France
\and European Southern Observatory, Karl-Schwarzschild-Str. 2, D-85748 Garching bei M\"unchen, Germany
\and Max-Planck-Institut f\"ur Astronomie, K\"onigstuhl 17, D-69117 Heidelberg, Germany
\and Departamento de F{\'\i}sica e Astronomia, Faculdade de Ci\^encias da Universidade do Porto, 4150-762 Porto, Portugal
}

\date{Received XXX; accepted XXX}
\abstract
{

We present the discovery of four new long-period planets within the HARPS high-precision sample: \object{HD137388}b ($M\sin{i}$ = 0.22 $M_J$), \object{HD204941}b ($M\sin{i}$ = 0.27 $M_J$), \object{HD7199}b ($M\sin{i}$ = 0.29 $M_J$), \object{HD7449}b ($M\sin{i}$ = 1.04 $M_J$). A long-period companion, probably a second planet, is also found orbiting HD7449. Planets around HD137388, HD204941, and HD7199 have rather low eccentricities (less than 0.4) relative to the 0.82 eccentricity of HD7449b. {All these planets were discovered even though their hosting stars have clear signs of activity. Solar-like magnetic cycles, characterized by long-term activity variations, can be seen for HD137388, HD204941 and HD7199, whereas the measurements of HD7449 reveal a short-term activity variation, most probably induced by magnetic features on the stellar surface. We confirm that magnetic cycles induce a long-term radial velocity variation and propose a method to reduce considerably the associated noise.} The procedure consists of fitting the activity index and applying the same solution to the radial velocities because a linear correlation between the activity index and the radial velocity is found. Tested on HD137388, HD204941, and HD7199, this correction reduces considerably the stellar noise induced by magnetic cycles and allows us to derive precisely the orbital parameters of planetary companions.

 

}

\authorrunning{Dumusque et al.}
\titlerunning{Planetary systems around stars with activity variation}
\maketitle

\section{Introduction} \label{sect:1}

{The radial velocity (RV) method has been the most successful technique until now in discovering new planets\footnote{see The Extrasolar Planets Encyclopaedia, http://exoplanet.eu}. The most effective RV machine at present, the HARPS spectrograph, now reaches the sub-meter-per-second precision level \citep[][]{Mayor-2009b,Lovis-2006}. Although the instrumental precision is excellent, the technique has some problems in identifying very low mass planets far away from their stars because we begin then to be sensitive to the noise caused by the stars themselves. A good understanding of the stellar noise is required if we wish to a) find Earth-like planets in the habitable regions of solar-type stars using the RV technique and b) confirm the 1200 planet candidates announced by the Kepler mission team \citep[][except for some systems showing transit timing variations]{Borucki-2011}.}

The RV technique is an indirect method that does not allow us to directly detect a planet: it measures the (gravitational) stellar wobble induced by the planetary companion orbiting its host star. Thus, the technique is sensitive not only to eventual companions, but also to stellar noise. At the sub-meter-per-second precision level, we begin to be affected by oscillation, granulation, and activity noises \citep[][]{Dumusque-2011a, Dumusque-2011b, Boisse-2011,Arentoft-2008,Huelamo-2008,Santos-2004a, Queloz-2001}. In this paper, we are interested in these different types of activity noise.

It has already been observed and demonstrated that magnetic features such as spots or plages on the stellar surface induce a RV signature when the star is rotating. Owing to rotation, one part of the star will be blueshifted whereas the other part will be redshifted. This effect is symmetric and the average RV will be zero. Active regions, which have a different flux than the solar surface, will break this balance and introduce a variation in the RVs coupled with rotation {\citep[][]{Boisse-2011,Queloz-2009,Henry-2002,Queloz-2001,Saar-1998,Saar-1997,Baliunas-1983,Vaughan-1981}}. We note that since this noise is proportional to the rotational velocity of the star ($v\sin{i}$), for similar magnetic features, the RV amplitude of the effect will be more significant for rapid rotators. Active regions contain spots, which are cooler than the average stellar surface, as well as plages, which are hotter. Therefore, the noise induced by active regions will usually be compensate for, but not entirely because of variation in the surface ratio of spots to plages \citep[e.g.][]{Chapman-2001}. The RV rms variation caused by spots and plages at the maximum of the Sun's activity is estimated to 48 cm\,s$^{-1}$ and 44 cm\,s$^{-1}$, respectively \citep[][]{Meunier-2010a}. We find a similar value for spots, 51 cm\,s$^{-1}$, when simulating the Sun's activity \citep[][]{Dumusque-2011b}. The total effect of spots plus plages reported by \citet{Meunier-2010a} for the same activity level is 42 cm\,s$^{-1}$, which shows the compensation effects between the two types of active regions. This noise is induced by activity coupled with rotation and therefore appears on short-period timescales. Below we refer to this as short-term activity noise.

{Concerning magnetic cycles, the idea that long-period stellar activity variation can perturb RVs dates back to \citet{Campbell-1988} and \citet{Dravins-1985}. Variation in the activity of thousand of FGK dwarfs, using the Mount-Wilson S-index and the log(R'$_{HK}$) index have been studied for twenty years \citep[][]{Hall-2007,Baliunas-1995,Duncan-1991}. However, no straightforward correlation between the long-term activity variation and the RV has been indentified until now \citep[][]{Santos-2010b,Isaacson-2010,Wright-2005,Wright-2004,Paulson-2004,Paulson-2002}. In this paper, we show that this correlation exists and the previous non-detections can easily be explained by the precision needed and the long-term RV follow up required to detect the effect  of magnetic cycles on RVs. According to \citet{Dravins-1985} and \citet{Meunier-2010a}, the inhibition of convection in active regions could be responsible for the correlation.} Because solar-like stars possess an external convective envelope, granulation will appear on the surface. Hot granules of plasma come to the surface, whereas cooled plasma flows into the interior, surrounding the hot granules and forming B\'enard cell patterns. Hot granules occupy a larger area than downward flows, which are less luminous. Because of the brightness-velocity correlation of the solar granulation \citep[][]{Dravins-1986,Kaisig-1982,Beckers-1978}, the surface will appear blueshifted in the presence of granulation. In active regions, the convection is greatly reduced \citep[][]{Meunier-2010a,Gray-1992,Brandt-1990,Livingston-1982,Dravins-1982} leading to a redshift of the spectrum relative to the average solar spectrum. Solar-like magnetic cycles are characterized by an increasing filling factor of magnetic regions when the activity level rises. {The total star convection is therefore lower and the star will appear redder (positive velocity) during its high-activity phase}. A positive correlation between the RVs and the activity level is then suspected \citep[][]{Dumusque-2010,Meunier-2010a}. {This explanation is the most commonly given, although some other physical processes could be responsible for this long-term RV-activity correlation, for example the variation in surface flows proposed by \citet{Makarov-2010}.}

We present here our detection of four new planet systems that have host stars with clear signs of activity. Three stars display solar-like magnetic cycles (long-term activity), whereas one star has short-term activity noise, related to magnetic features on its surface. In Sects. \ref{sect:2} and \ref{sect:3}, informations about the observations made and the stellar parameters of the hosting stars are given, and in Sect. \ref{sect:5}, planet properties for each star are derived. Finally Sect. \ref{sect:6} discusses our results. In addition, Table 4, 5, 6, and 7, only available in the online version and at the CDS, contains the RVs and the activity index for each star.

\section{Sample and observations} \label{sect:2}
	
The observations were carried out as part of the HARPS high-precision programs, gathering together a total of 451 stars. The majority of measurements were obtained during a guaranteed time observation \citep[GTO, PI: M. Mayor,][]{Mayor-2003b}. However some additional measurements have been carried out under an ESO large program (PI: S. Udry) to define precisely the orbital solutions.

RVs were been derived using version 3.5 of the HARPS Data Reducing System (DRS). For each spectrum, the present pipeline returns several parameters of the cross correlation function (CCF) : the RV, the full width at half maximum (FWHM), the BISsector span of the CCF inverse slope \citep[BIS span,][]{Queloz-2001} and the ratio of the CCF minimum to the continuum (CONTRAST). The activity level, expressed in the Mount Wilson S-index \citep[][]{Vaughan-1978} and in the chromospheric emission ratio, log($R'_{HK}$) \citep[][]{Noyes-1984}, is given as well.

All the measurements were derived using the simultaneous thorium calibration technique and exposure times were set to 15 minutes to enable us to average out stellar jitter induce by oscillation modes \citep[][]{Dumusque-2011a,Santos-2004a}. This allows us to reach a precision of higher than 1\,m\,s$^{-1}$ on the derived RVs. The averaged signal-to-noise ratio (SNR) at 550 nm varies between 120 and 137, depending on the star.

\section{Stellar parameters} \label{sect:3}

Table \ref{tab:1} regroups the stellar parameters for each star analyzed in the present paper.

The spectral type, $V$, $B-V$, and the parallax with the derived distance are taken from the Hipparcos catalog \citep[][]{ESA-1997}. The parameters $M_V$, $L$, and $R$ are derived from the above values, using the bolometric correction from \citet{Flower-1996} \citep[see][for the correct polynomial index]{Torres-2010}, and the $T_{\mathrm{eff}}$ obtained from the spectroscopy.

The spectroscopic analysis of the four stars was conducted assuming local thermodynamical equilibrium and using Kurucz atmosphere models. The equivalent width of the spectral lines were computed via the ARES code \citep[][]{Sousa-2007}. This analysis enables us to derive $T_{\mathrm{eff}}$, log $g$, and $[$Fe/H$]$. The list of spectral lines used for the analysis, as well as more details about the method can be found in \citet{Sousa-2008} and \citet{Santos-2004b}

The activity index, $\log{R'_{HK}}$, was extracted from individual HARPS spectra using a method similar to the one described in \citet{Santos-2000a} (Lovis et al., in prep.). Table 1 shows for each star the mean level of the activity index, as well as its full variation amplitude, $\Delta$(log($R'_{HK}$)), and its standard deviation, $\sigma$(log\,$R'_{HK}$). {For the three early-K dwarfs, the mean activity level is low but the variation in the activity index is above 0.1 dex owing to the solar-like magnetic cycles (see Sec. \ref{sect:5}).} For the late-F dwarf, HD7449, the level of activity is high without any long-term variation. This agrees with \citet{Baliunas-1995}, where the authors find that F dwarfs have very low or non-existent long-term activity variation. The rotation period and the age for each star is estimated from the activity level and the color index, as described in \citet{Mamajek-2008}, and \citet{Noyes-1984}.
%
%
\begin{table*} 
	\begin{center}
		\caption{Stellar parameters for the stars analyzed in the present paper. The value log(R'$_{HK}$) refers to the mean value of the activity level, whereas $\Delta$(log\,(R'$_{HK}$)) and $\sigma$(log\,(R'$_{HK}$)) give the amplitude and the rms of the activity index variation, respectively.}  \label{tab:1}
		\begin{tabular}{ccccc}
			\hline
			\hline Parameters & HD7199 & HD7449 & HD137388 & HD204941\\
			\hline
			Spectral type & K0IV/V & F8V & K0/K1V & K1/K2V\\
			$V$ 						& 8.06 & 7.50 & 8.71 & 8.45\\
			$B-V$ 					& 0.849 & 0.575 & 0.891 & 0.878\\
			Parallax [mas] 				& 27.87 $\pm$ 0.69 & 25.96 $\pm$ 0.77 & 26.01 $\pm$ 1.88 & 37.10 $\pm$ 1.26\\
			Distance [pc] 				& 36 $\pm$ 1 & 39 $\pm$ 1 & 38 $\pm$ 3 & 27 $\pm$ 1\\
			$M_v$ 					& 5.29 & 4.57 & 5.79 & 6.30\\
			$L\,[L_{\odot}]$				& 0.70 & 1.21 & 0.46 & 0.31\\
			$R\,[R_{\odot}]$	 		& 0.96 & 1.01 & 0.82 & 0.72\\
			$M\,[M_{\odot}]$		 	& 0.89 & 1.05 & 0.86 & 0.74\\
			$T_{\mathrm{eff}}$ [K] 		& 5386 $\pm$ 45 & 6024 $\pm$ 13 & 5240 $\pm$ 53 & 5056 $\pm$ 52\\
			log $g$ 					& 4.34 $\pm$ 0.08 & 4.51 $\pm$ 0.03 & 4.42 $\pm$ 0.11 & 4.48 $\pm$ 0.09\\
			$[$Fe/H$]$	 			& 0.28 $\pm$ 0.03 & -0.11 $\pm$ 0.01 & 0.18 $\pm$ 0.03 & -0.19 $\pm$ 0.03\\
			log($R'_{HK}$)				& -4.95 & -4.85 & -4.90 & -4.94 \\
			$\Delta$(log($R'_{HK}$))	 	& 0.31 & 0.06 & 0.22 & 0.13\\
			$\sigma$(log($R'_{HK}$)) 	& 0.083 & 0.014 & 0.053 & 0.032\\
			$P_{rot}$ 					& 42.89 $\pm$ 4.66 & 13.30 $\pm$ 2.56 & 42.44 $\pm$ 4.96 & 43.6 $\pm$ 4.88\\
			Age\,[Gyr] 					& 6.79 $\pm$ 0.78 & 2.10 $\pm$ 0.24 & 6.45 $\pm$ 0.74 & 6.67 $\pm$ 0.77\\
			\hline
		\end{tabular} 
	\end{center}
\end{table*}
%


\section{Radial velocity data and orbital solution} \label{sect:5}

{We present our detections of four new planetary systems and the characteristics of each companion. The orbital parameters were derived by adding quadratically 1 and 5\,m\,s$^{-1}$ to the RV noise estimations of HARPS and CORALIE, respectively. This can be justified because only the photon noise, generally the most dominant noise, the calibration noise ($\sim$ 0.30 m.s$^{-1}$ for HARPS) and the instrumental drift ($\sim$ 0.10 m.s$^{-1}$ for HARPS) are taken into account in the RV noise estimation. The guiding noise, depending greatly on the weather conditions, as well as the stellar noise, which is only approximately known given the star spectral type and activity level \citep[][]{Dumusque-2011a,Dumusque-2011b}, are not included. Therefore adding quadratically the RV rms of the quietest stars in the HARPS and CORALIE sample (1\,m\,s$^{-1}$ and 5\,m\,s$^{-1}$, respectively) avoids large and unjustified differences in the individual weights ($w_i$= 1/$\sigma_i^2$) used during the $\chi^2$-minimization process.}


\subsection{HD7199} \label{sect:5.1}

From BJD =  2452946 (November 3,  2003) to BJD = 2455524 (November 24, 2010), we obtained 88 measurements of HD7199 with an average SNR of 137 at 550 nm (the minimum value is 55). {The mean RV uncertainty returned by the DRS}, 0.60 m\,s$^{-1}$, is dominated by the photon noise.

\begin{figure*}
\begin{center}
\includegraphics[width=8cm]{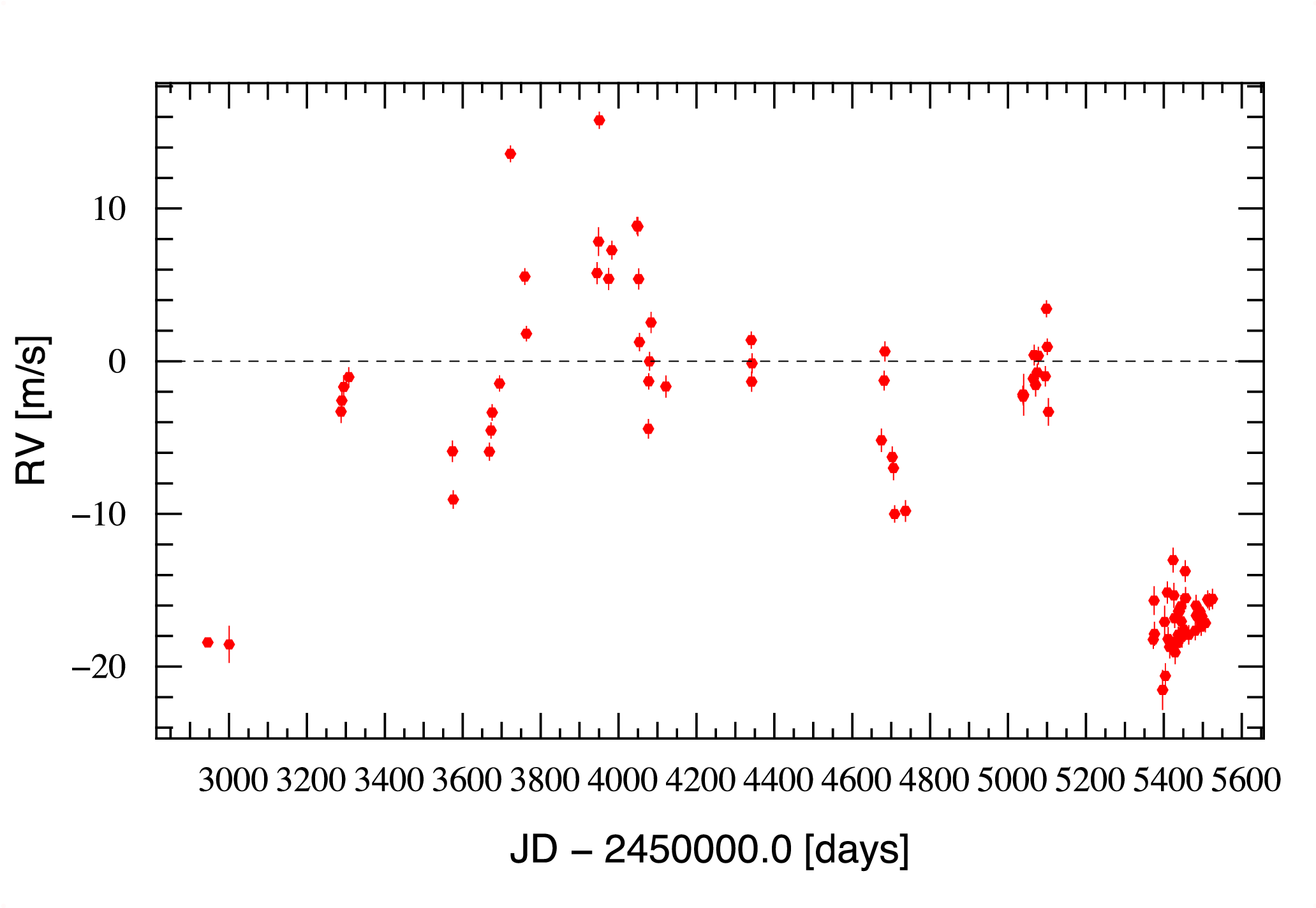}
\includegraphics[width=8cm]{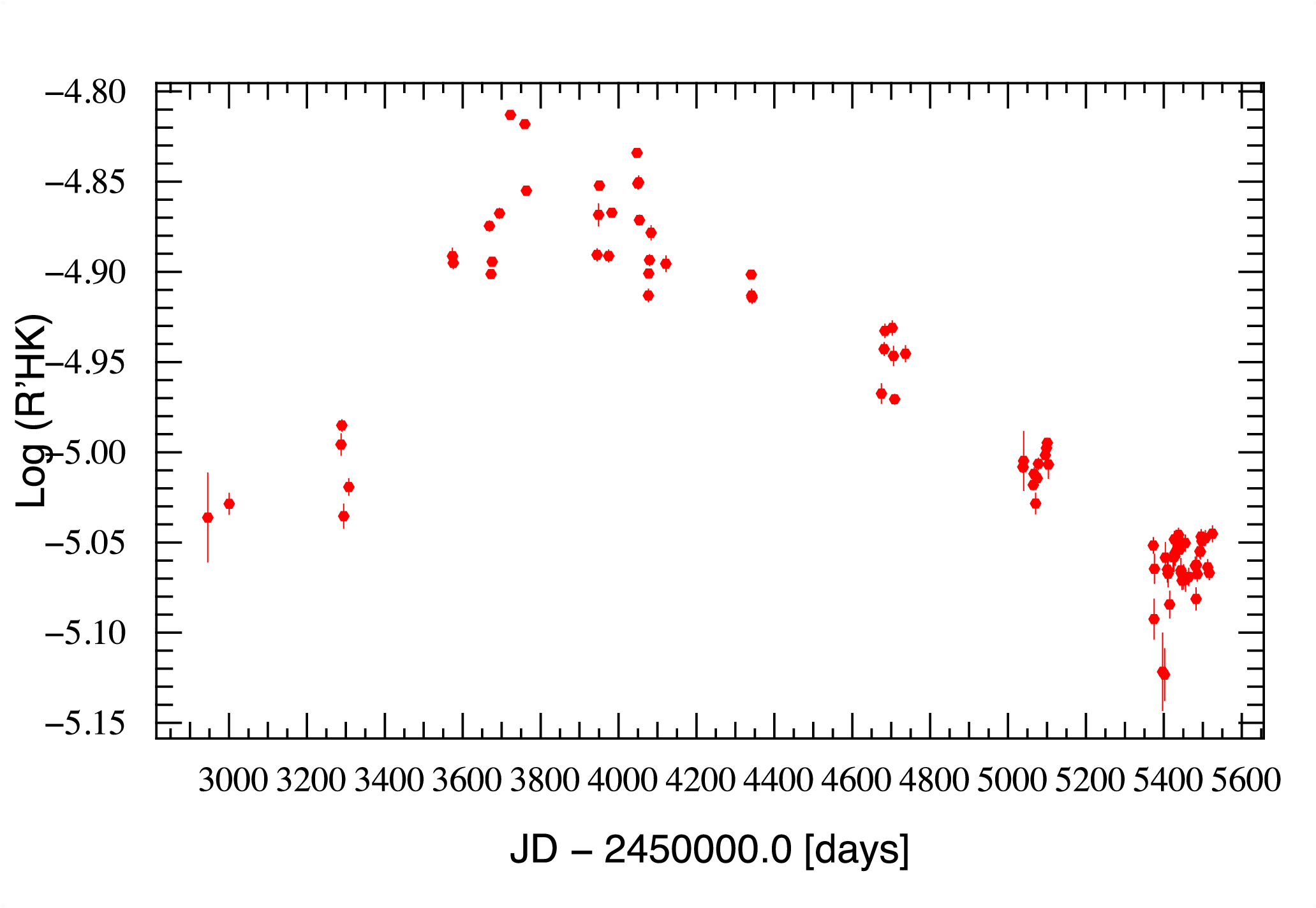}
\includegraphics[width=8cm]{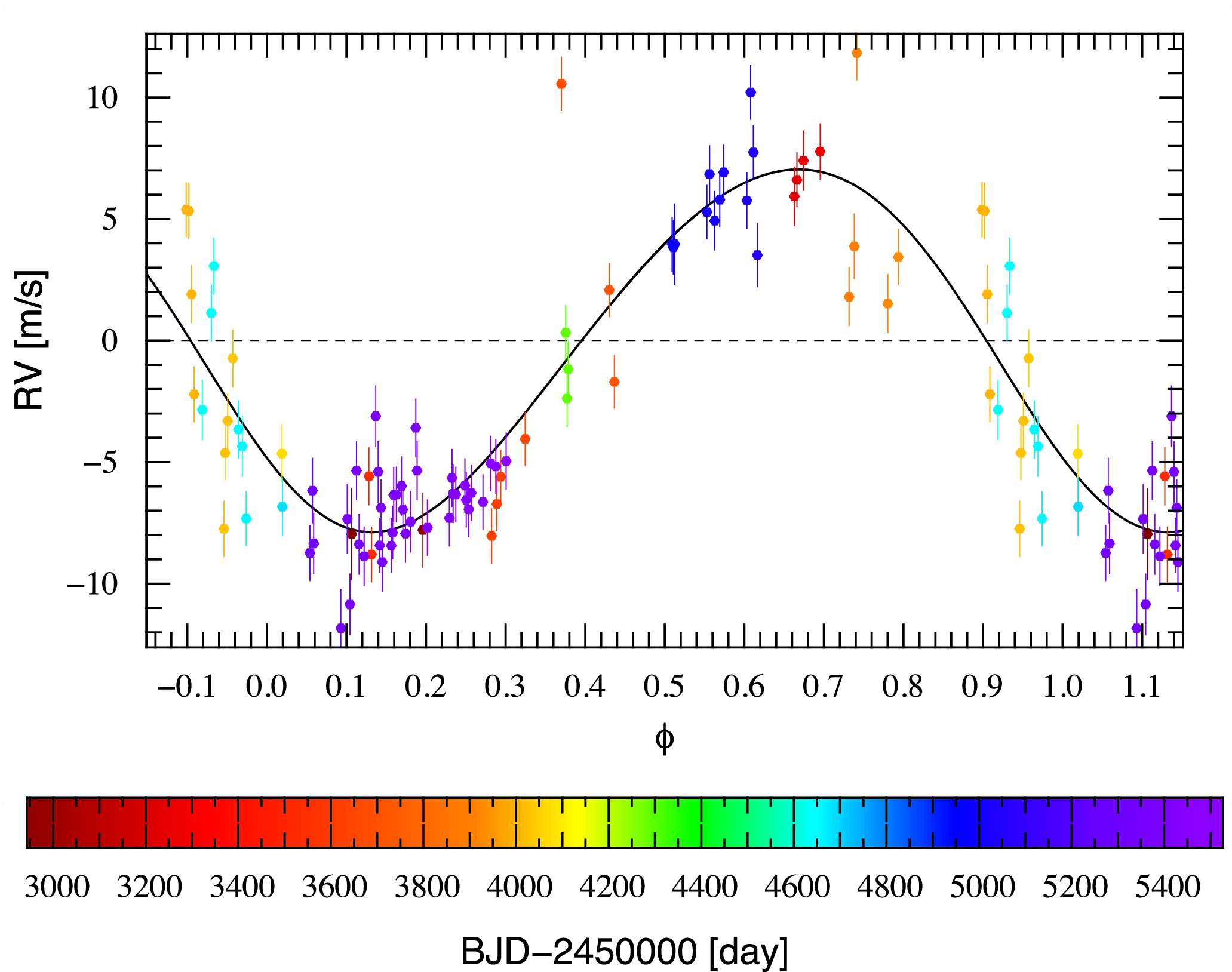}
\includegraphics[width=8cm]{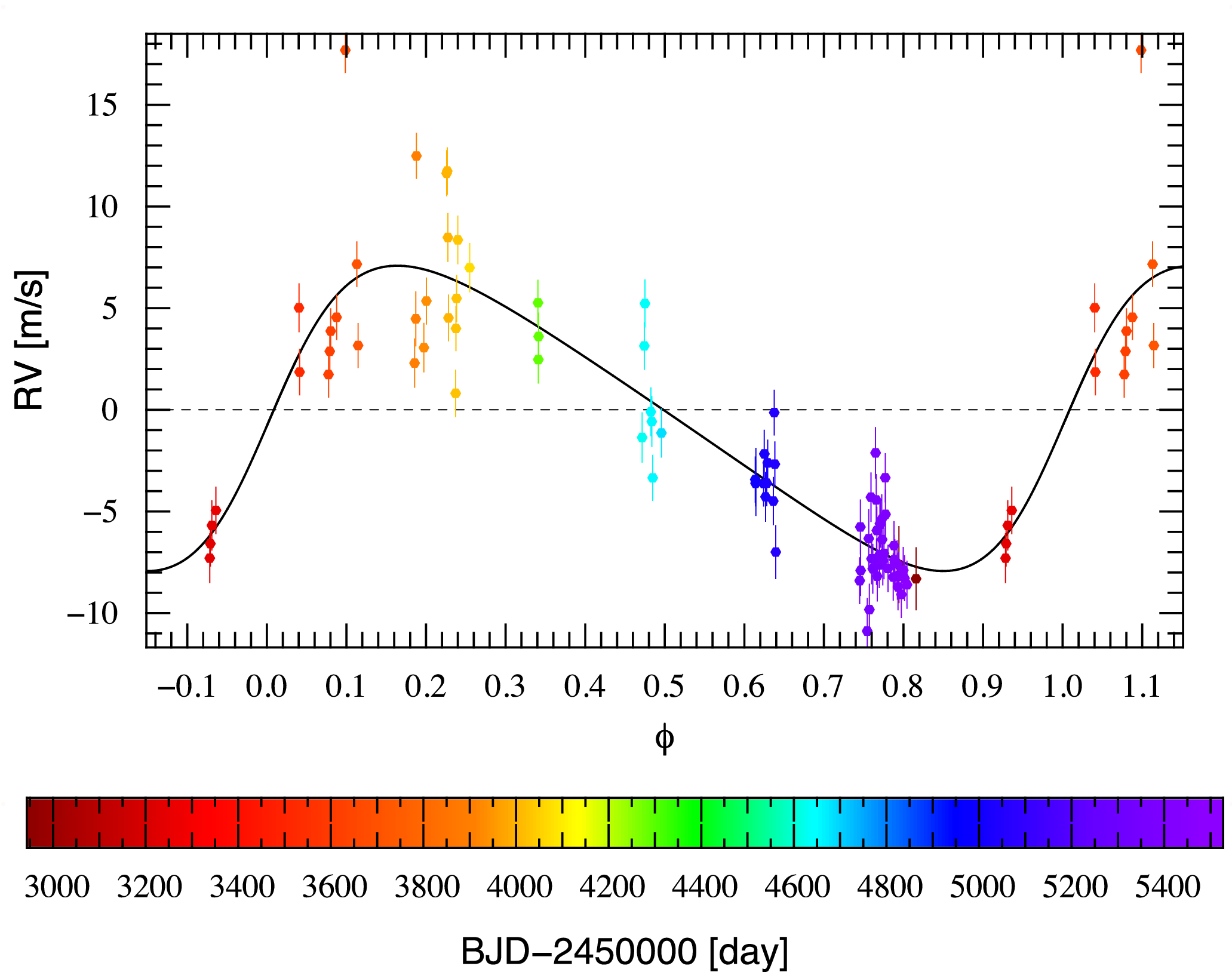}
\caption{All the graphs are for HD7199. \emph{Top panel: }RVs and activity index, log(R'$_{HK}$), as a function of time. \emph{Lower panel: }RVs folded in phase for the two Keplerian fitted. On the left the real planet and on the right the RV variation induced by the magnetic cycle, which is similar to the activity index variation. Thus, we see clearly that magnetic cycles induce a RV variation.}
\label{fig:0}
\end{center}
\end{figure*}

\begin{figure*}
\begin{center}
\includegraphics[width=8cm]{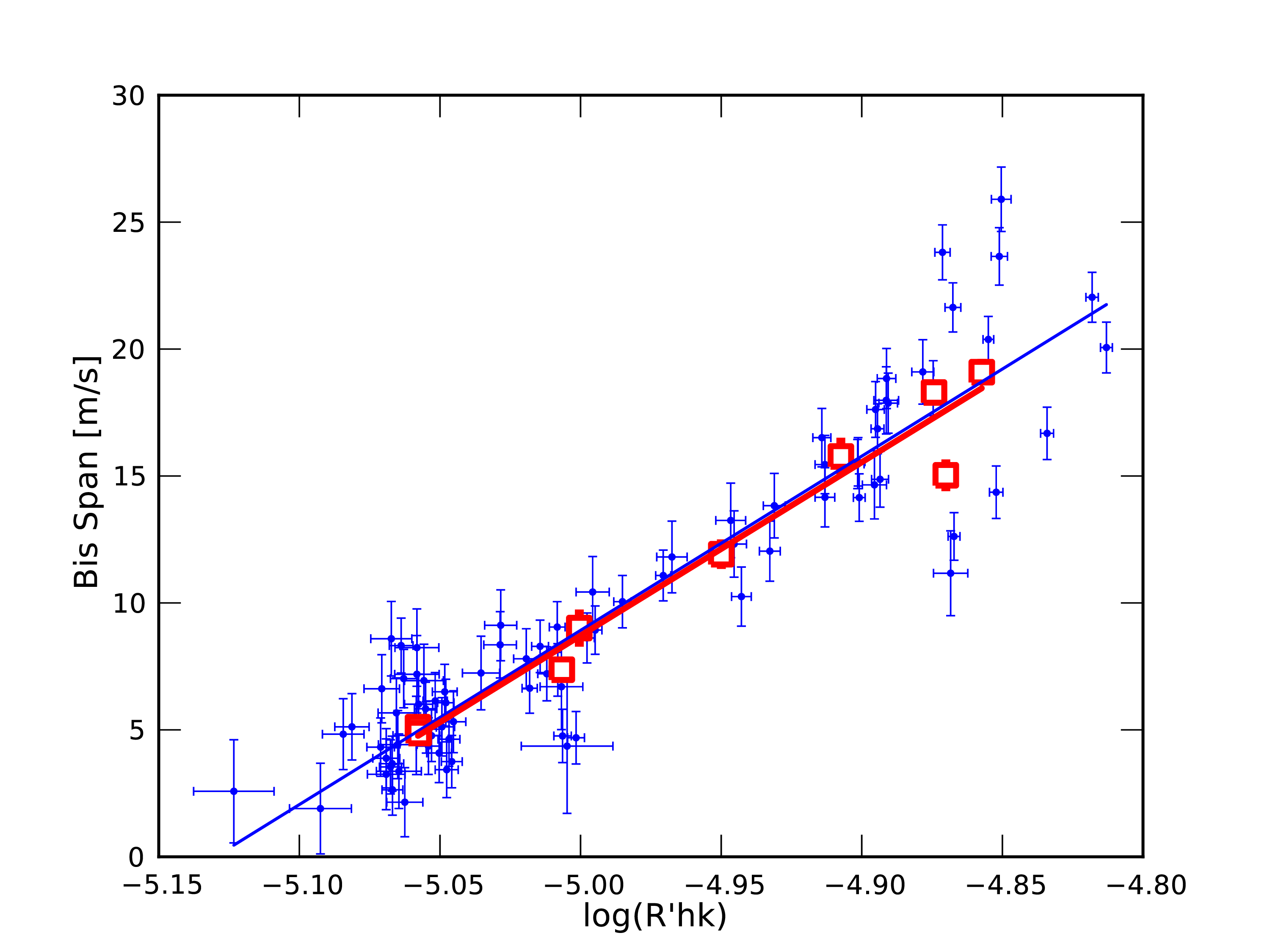}
\includegraphics[width=8cm]{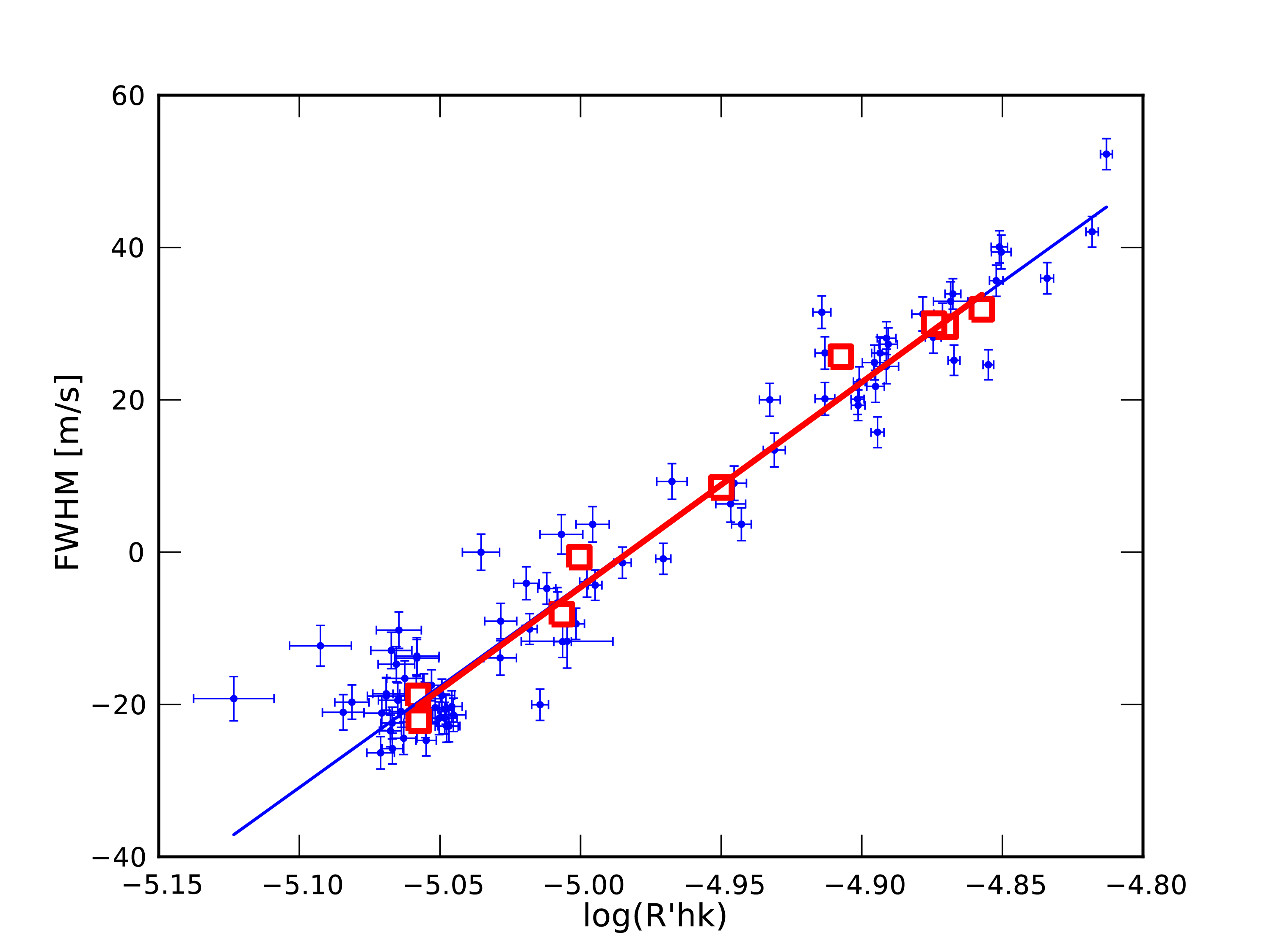}
\includegraphics[width=8cm]{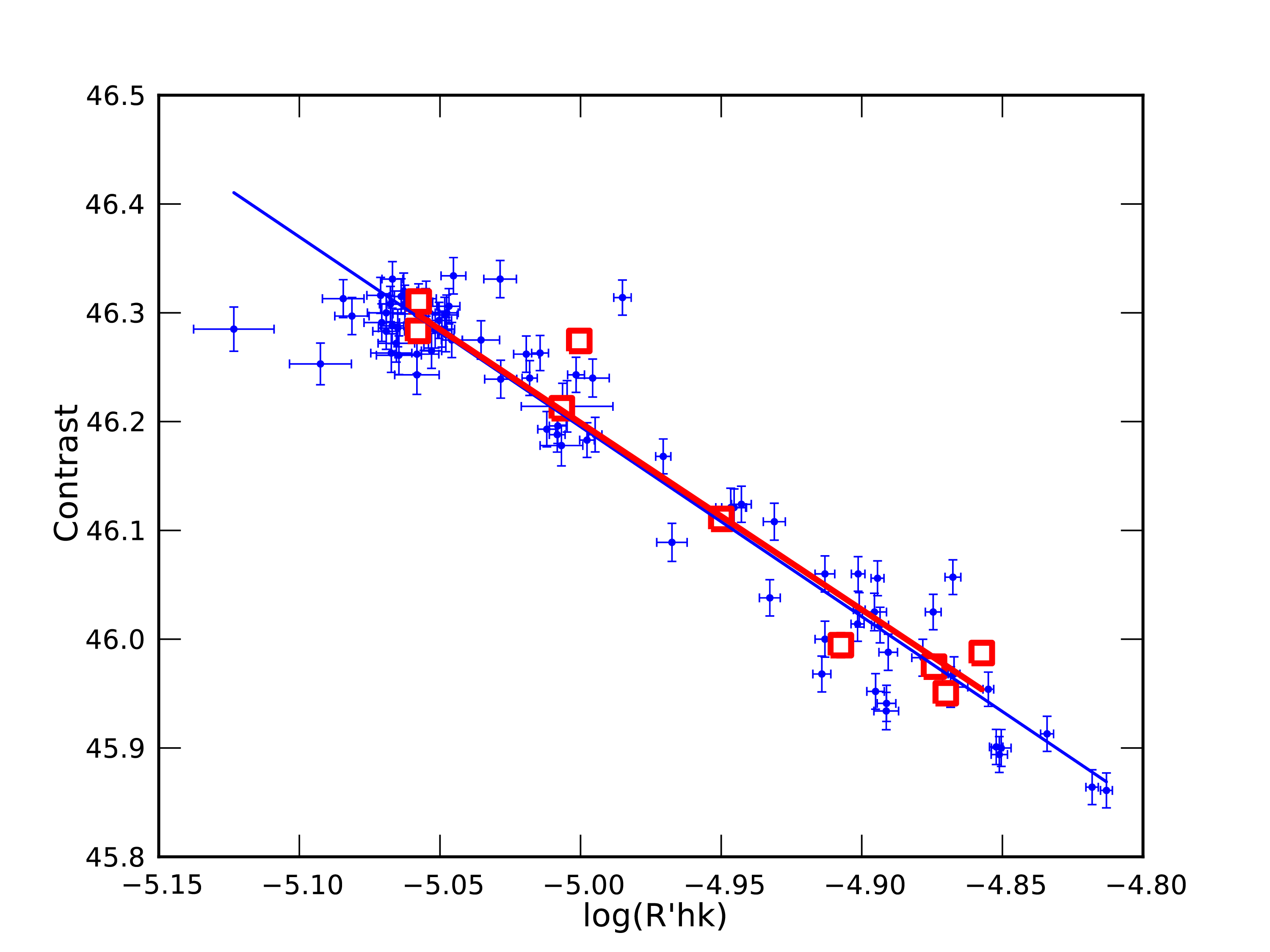}
\includegraphics[width=8cm]{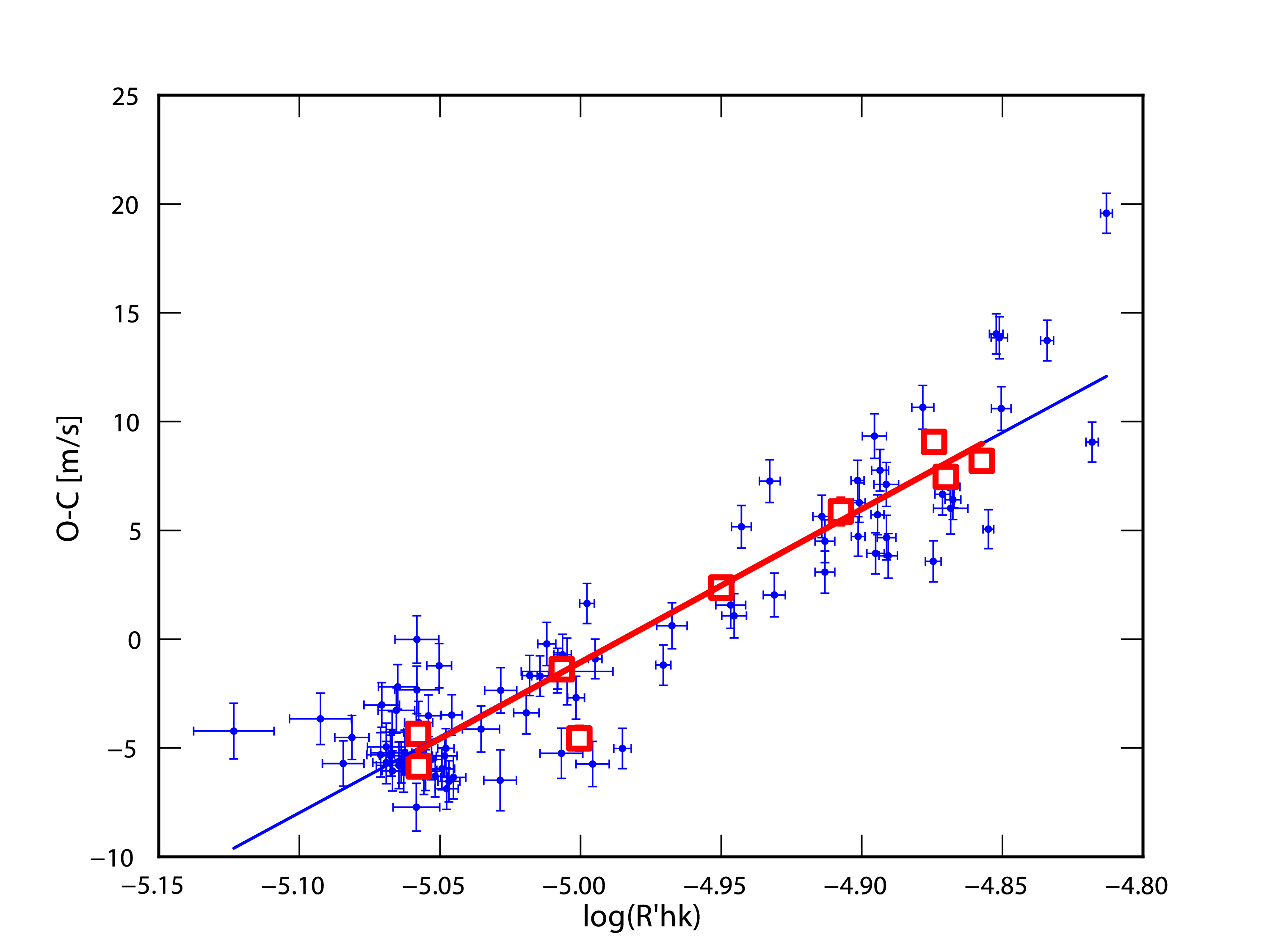}
\caption{Correlation between the CCF parameters and the activity index for HD7199. The RV residuals (O-C) correspond to the raw RVs after removing the detected planet. Small points correspond to all the measurements, whereas big squares represent the same data binned over 3 months to average out short-term activity.}
\label{fig:2}
\end{center}
\end{figure*}

Looking at the activity index log(R'$_{HK}$) as a function of time, we see a clear variation due to a solar-like magnetic cycle (Fig. \ref{fig:0} \emph{top right}). {This cycle is incomplete, therefore only a minimum amplitude of 0.25 dex and a minimum period of seven years can be extracted. Given these minimum estimations, this magnetic cycle is very similar to the solar one.} To investigate whether this activity variation can have some influence on the CCF parameters, we calculated the correlation between them and the activity index (Fig. \ref{fig:2}). {In Fig. \ref{fig:2}, the red squares represent a binning in time, which allows us to reduce the effect of short-term activity. The small difference between the correlation obtained with the binned data and the raw ones implies that for this star, the log(R'$_{HK}$) variation induced by short-term activity is smaller than the one induced by the magnetic cycle. All the parameters, even the RV residuals after removing the discovered planet (see below), are very strongly correlated with activity, therefore they can be used to search for long-term activity phenomena. The raw RVs are obviously not correlated to activity, because the planet will introduce a RV variation without changing the activity level.}
\begin{figure}
\begin{center}
\includegraphics[width=8cm]{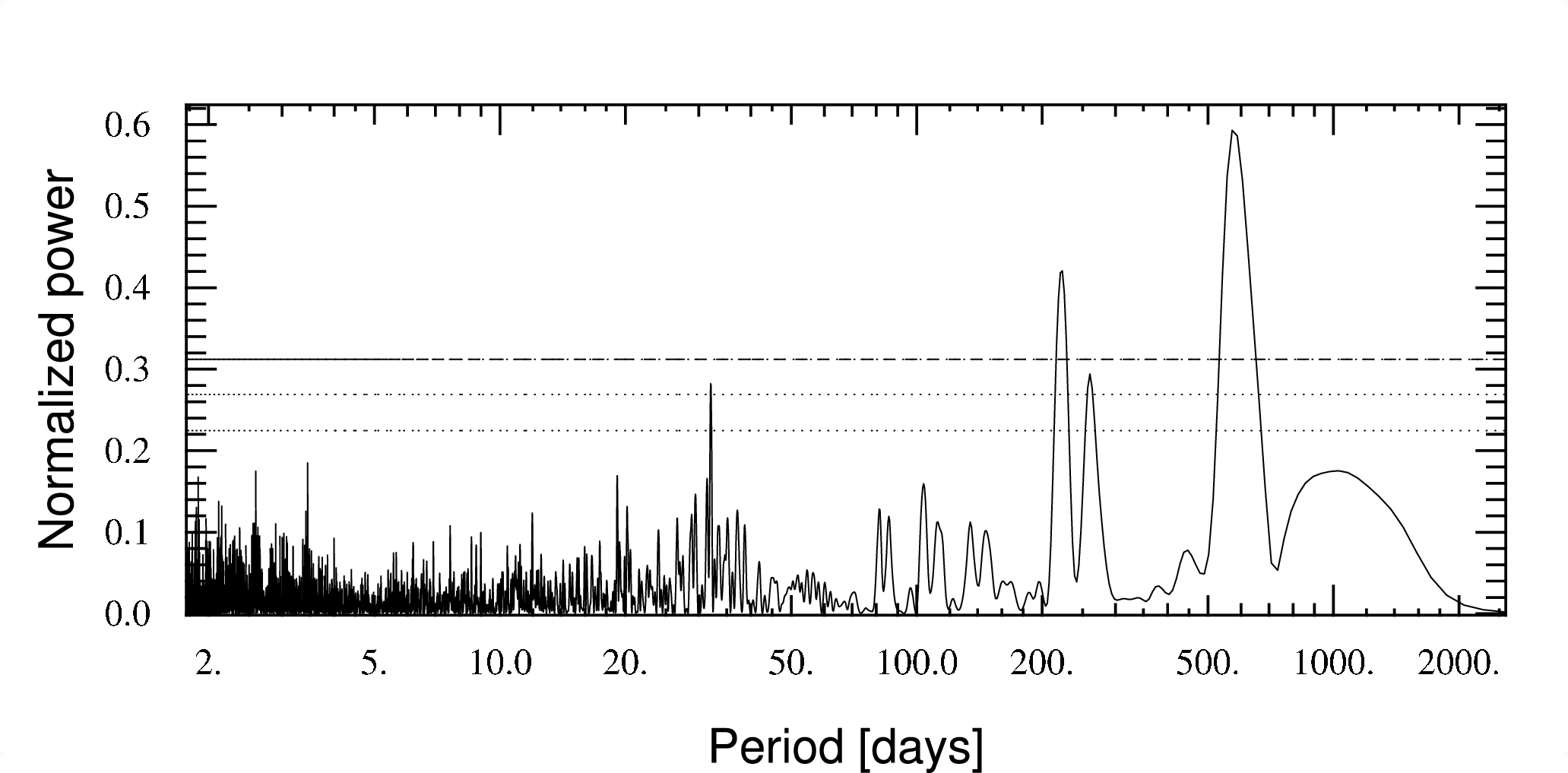}
\includegraphics[width=8cm]{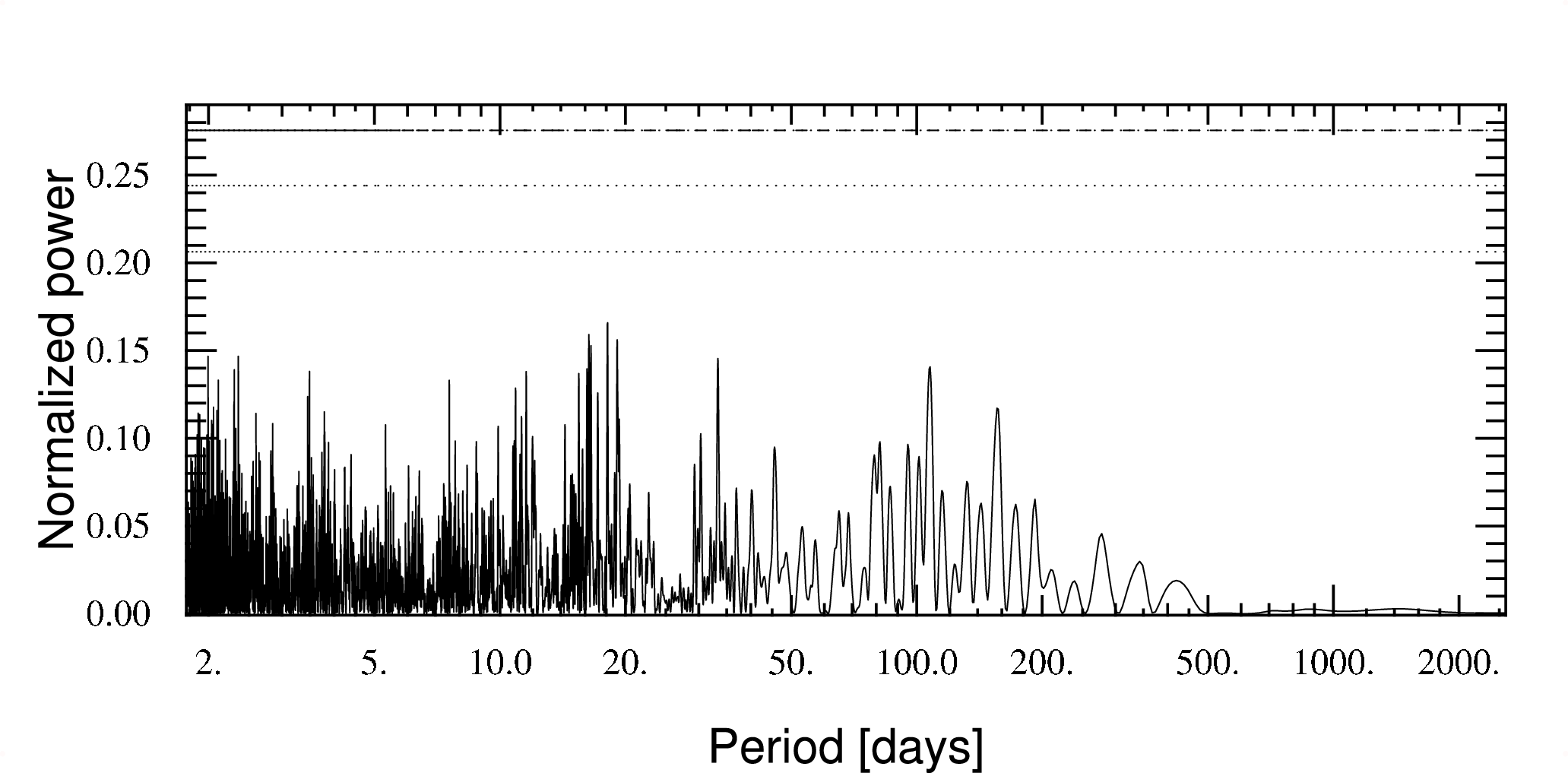}
\caption{Periodograms for HD7199 of the raw RVs after removing a quadratic drift (\emph{top}) and of the RV residuals after removing the planet and the effect of the magnetic cycle (\emph{bottom}). The horizontal lines correspond from top to bottom to a FAP of 0.1\,\%, 1\,\%, and 10\,\%.}
\label{fig:1}
\end{center}
\end{figure}

When fitting and subtracting a second order polynomial from the RVs, a very significant peak with a confidence level of 99.9\,\% appears near 600 days in the periodogram (Fig. \ref{fig:1}). This signal corresponds to a planet with a 615 day period and an eccentricity of 0.19. The raw RVs are therefore the superposition of a long-period planet signal and a RV variation induced by a magnetic cycle. {To subtract properly the magnetic cycle effect on the RVs, we first fit a Keplerian to the activity index. We then fit two Keplerians to the RVs, one for the planet and one for the activity for which only the amplitude is free to vary (the other parameters are locked to the values found when fitting the activity index).} We finally obtain an orbital solution with a 615 day period planet and an eccentricity of 0.19. The minimum mass of this object is 0.290 M$_J$ = 92 M${_{\oplus}}$. All the parameters of the fitted planet are listed in Table \ref{tab:2} and the fit can be seen in the lower panels of Fig. \ref{fig:0}. In the periodogram of the RV residuals (Fig. \ref{fig:1}), the highest peak is much lower than the 10\,\% false-alarm probability (FAP) excluding at this precision level the presence of a second companion.

The comparison between the activity index variation and the second Keplerian fit to the RVs (see Fig. \ref{fig:0} \emph{top right panel} and \emph{lower right panel}, respectively) shows that magnetic cycles induce a RV variation. Therefore, for long-period RV signals, activity variation must be studied very carefully to avoid false planet detection. In \citet{Meunier-2010a}, the authors claim that for the Sun, the RV amplitude induce by its magnetic cycle is of approximately 10\,m\,s$^{-1}$. We find a similar amplitude when looking at the long-term RV variation of HD7199.


The rms of the RV residuals is quite high for a K dwarf, 2.63\,m\,s$^{-1}$. One observational points at BJD = 2453722.6 seems to be an outlier in the raw RVs and is approximately 10\,m\,s$^{-1}$ far from the fitted orbit. When removing this point and applying the same fit, the RV residuals rms decreases to 2.06\,m\,s$^{-1}$, compatible with an activity level of log(R'$_{HK}$) = -4.8 (similar to the maximum solar activity). The SNR of this measurement is high (167) and looking at the La Silla meteo monitor for this night, the weather conditions were good. Therefore, we do not have any good explanation for this "bad" measurement.

\subsection{HD7449} \label{sect:5.2}

\begin{figure*}
\begin{center}
\includegraphics[width=8cm]{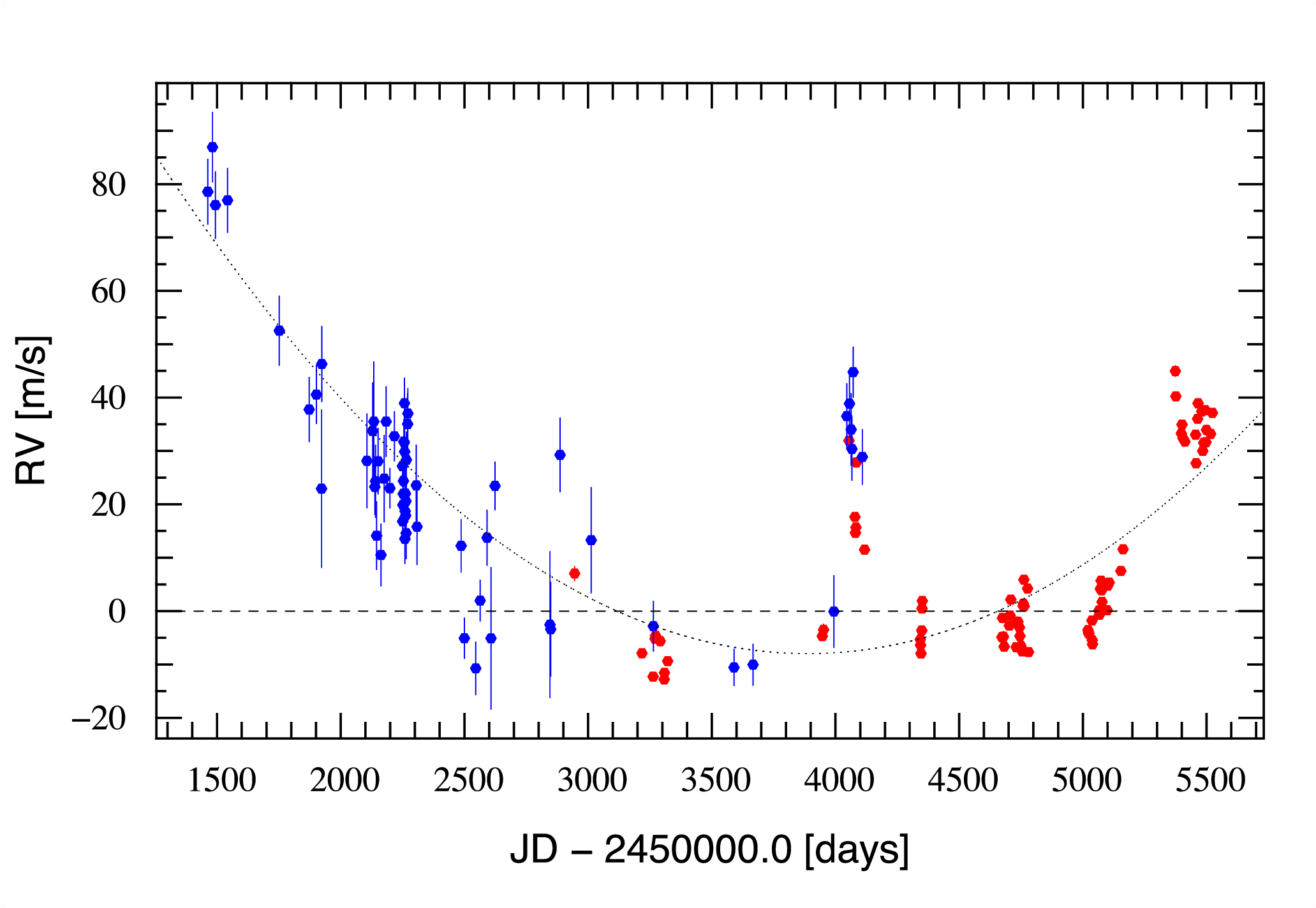}
\includegraphics[width=8cm]{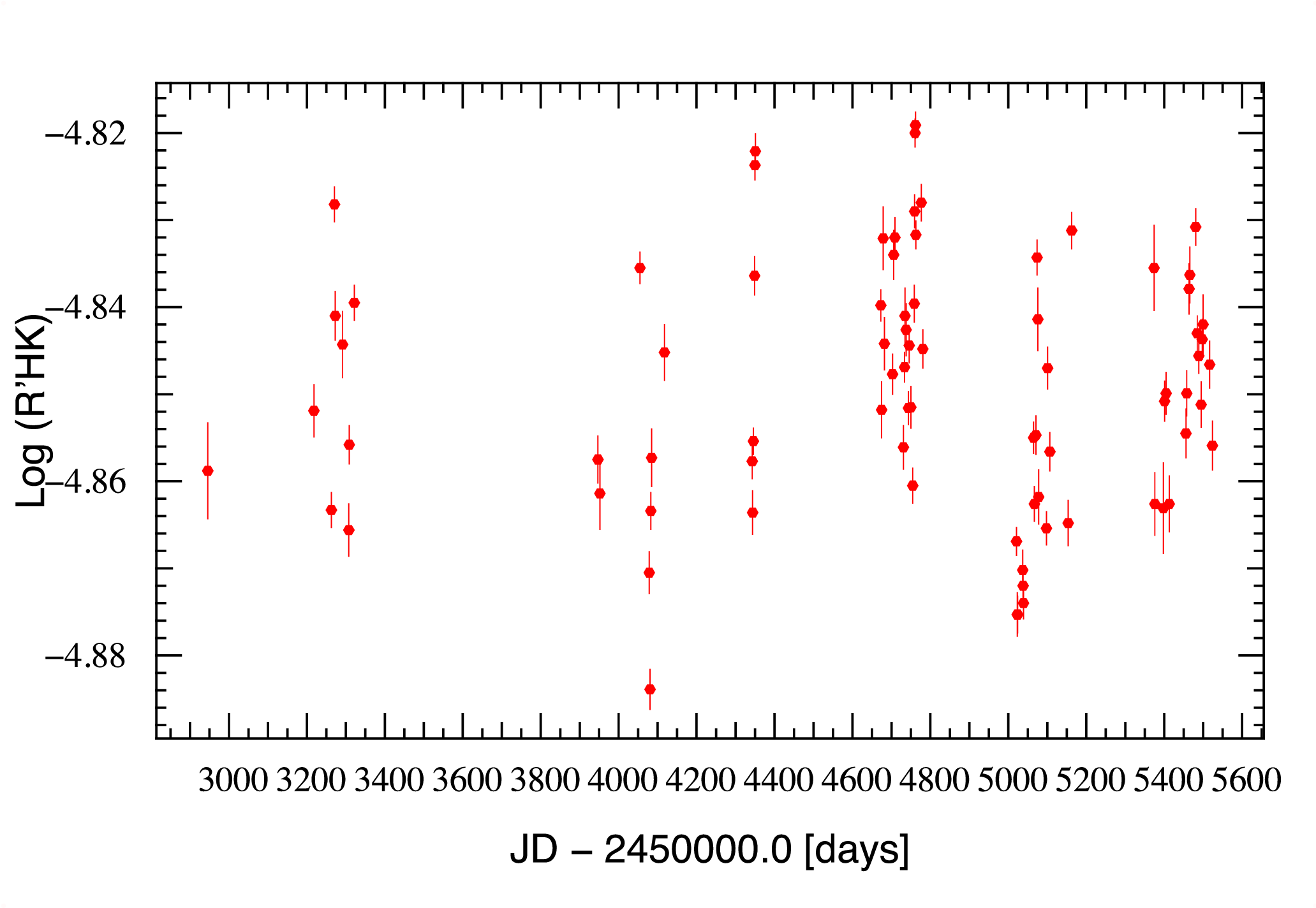}
\includegraphics[width=8cm]{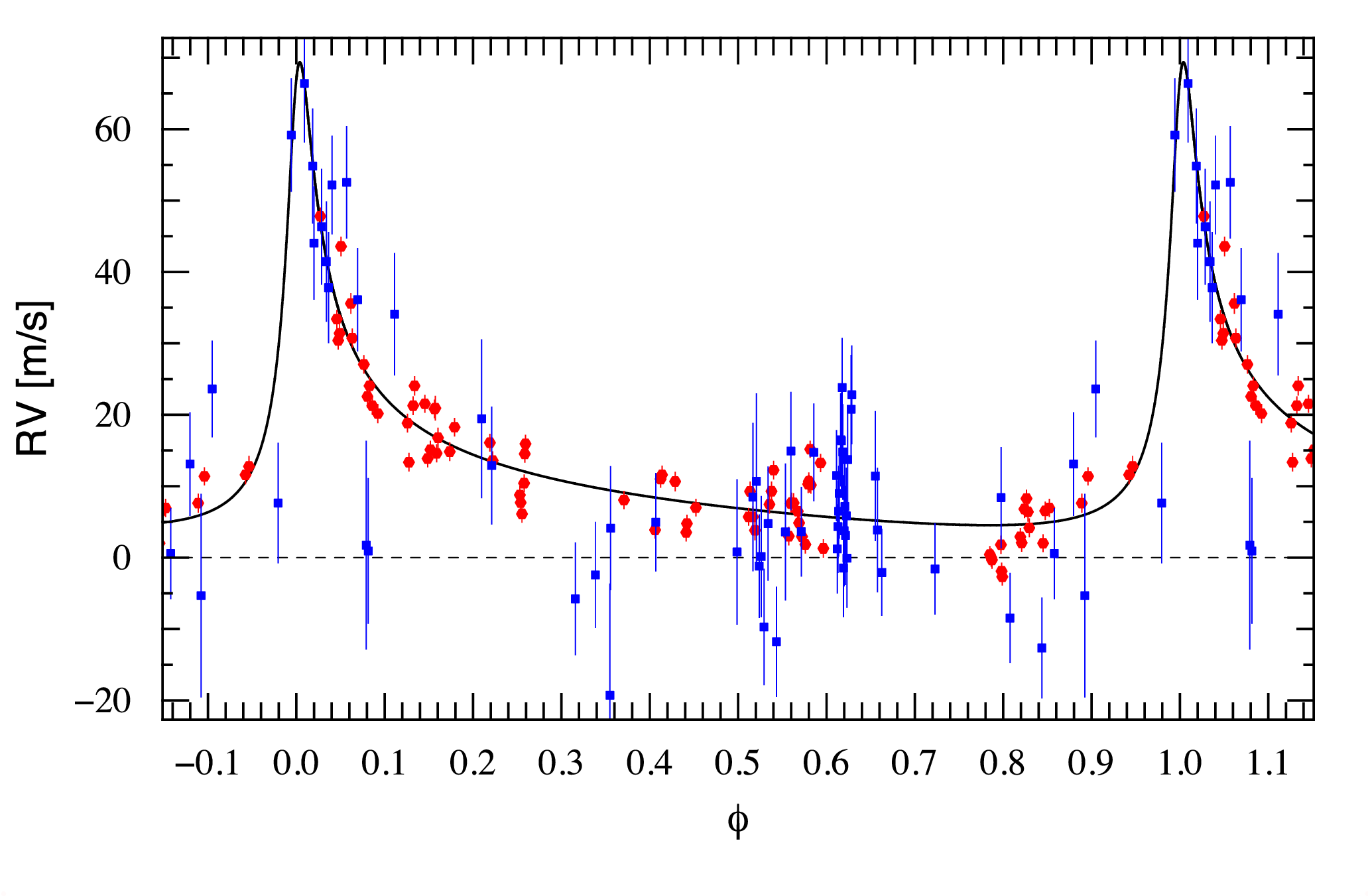}
\includegraphics[width=8cm]{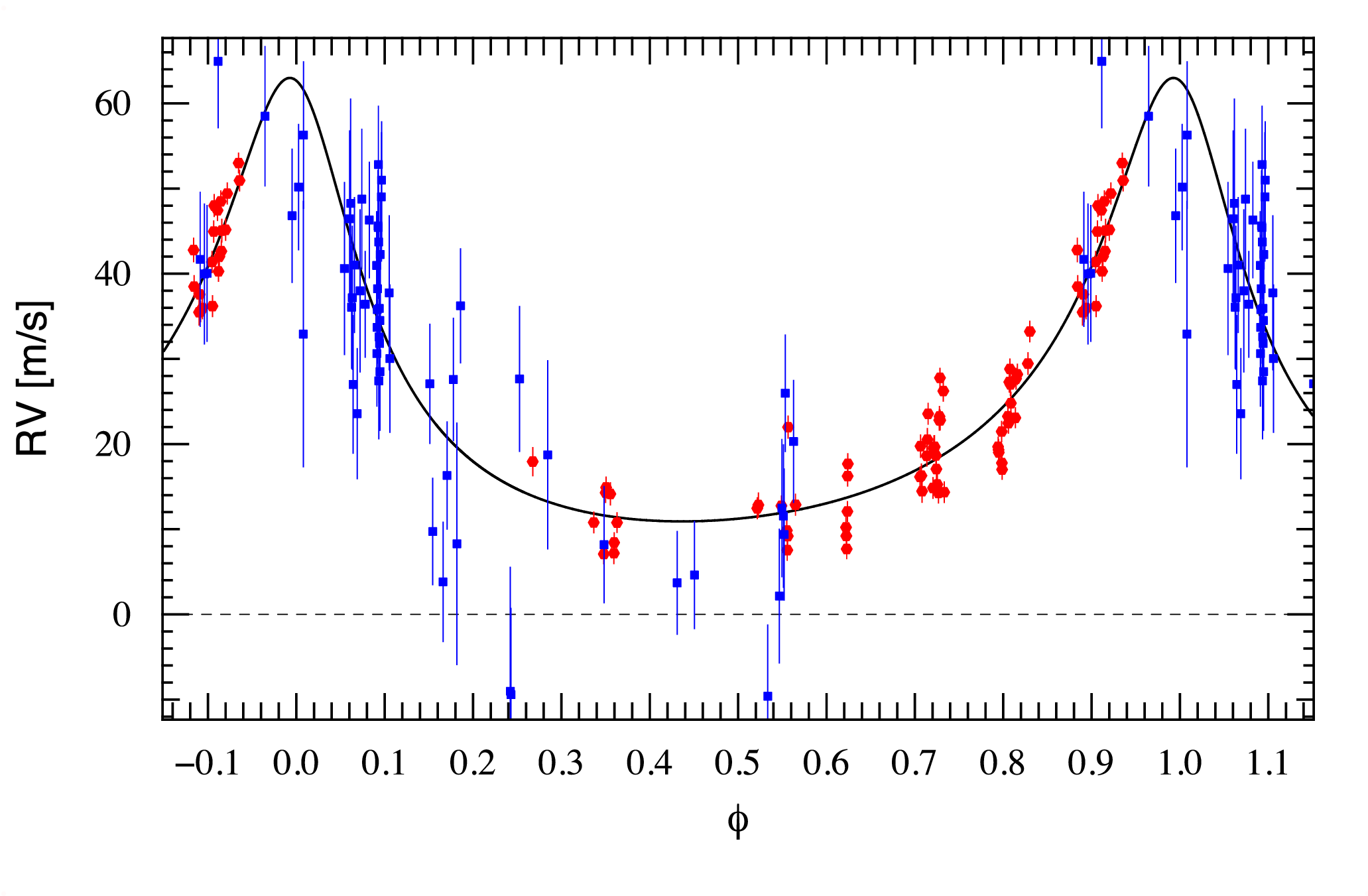}
\caption{All the graphs are for HD7449. \emph{Top panel: }RVs and the activity index, log(R'$_{HK}$), as a function of time. Blue points (dark) are for CORALIE and red points (gray) for HARPS. Note that the CORALIE pipeline does not return the activity index. \emph{Lower panel: } Orbital solution folded in phase for the best-fit solution of the short-period planet (\emph{left}) and the long-period one (\emph{right}).}
\label{fig:3}
\end{center}
\end{figure*}

Over 7.2 years, from BDJ = 2452946 (November 3,  2003) to BJD = 2455579 (January 17, 2011), HD7449 was observed 82 times with HARPS. RVs were extracted from high signal-to-noise ratio spectra ($<$SNR$>$ = 190 at 550 nm and minimum is 111) with a mean RV uncertainty of 0.69 m\,s$^{-1}$ (photon and calibration noise). In addition, HD7449 was observed 62 times with CORALIE over seven years (from BJD = 2451462 to BJD = 2454108), with $<$SNR$>$ = 54 at 550 nm and a mean RV uncertainty of 4.97 m\,s$^{-1}$. This star has a mean activity level $<log(R'_{HK})>$=-4.85, without any magnetic cycle as we can see in Fig. \ref{fig:3} \emph{top right panel}. This activity level, comparable to the one of the Sun at the maximum of its magnetic cycle, remains stable during the 7.2 years of the HARPS follow-up. 

Looking at the periodogram of the HARPS raw RVs, a very significant peak appears around 1300 days indicating the presence of a long-period planet. Moreover, the optimized orbital solution is provided by a two-component Keplerian fit to the data, one for the 1300-day period signal, and one for a long-term trend. When combining the data from HARPS and CORALIE, for a total span of 14 years, the long-term trend is confirmed and we find the optimal orbital solution for a planet with a 1275 day period, an eccentricity of 0.82, a reduced chi-square of 6.73, and a RV residual rms of 3.81\,m\,s$^{-1}$ (only calculated on HARPS data). We note that the high eccentricity of the planet is found because of the bump in RV near BDJ = 2454000 (see Fig. \ref{fig:3} \emph{top left panel}). This bump is well sampled by HARPS and CORALIE data, excluding possible instrumental errors. The long-term RV variation is not due to a magnetic cycle, because of the long-term activity stability, but could be induced by a second companion, for which the period has not been yet covered. However, a first estimation of its mass and period can be made. Table \ref{tab:2} gives the orbital solutions for the HD7449 system and Fig. \ref{fig:3} \emph{bottom} shows the corresponding fits. 

\begin{figure}
\begin{center}
\includegraphics[width=6cm]{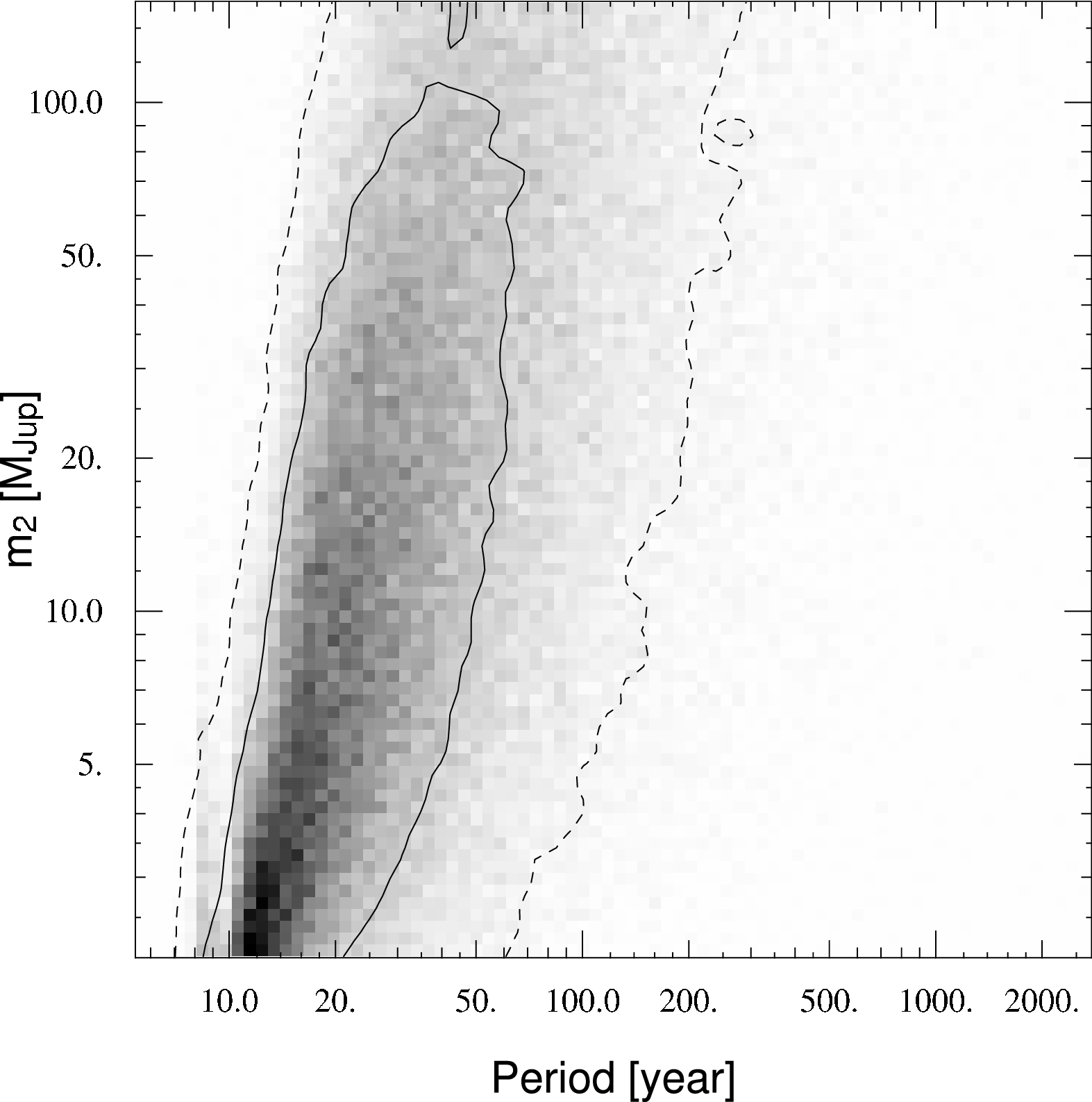}
\caption{Monte Carlo simulation of the outer companion of HD7449. The continuous contour corresponds to a level of 1\,$\sigma$ and the dashed one to 2\,$\sigma$.}
\label{fig:5}
\end{center}
\end{figure}

To determine the nature of this outer companion, we first considered whether a stellar companion could be responsible for this long-term variation. No candidates was found in the Washington Double Star catalogue, and nothing closer than 55 arcsec was observed in the 2MASS catalogue, which corresponds to a huge separation of 2150\,AU. We considered whether the system could be constrained by the Hipparcos astrometry, using the method described in \citet{Sahlmann-2011}. Neither the inner planet, nor the outer one could be detected by this technique. The low mass and the relatively short period of the inner planet explain this non-detection. For the outer companion, the orbital solution fitted (see Table \ref{tab:2}) gives a star minimal semi-major axis of $a_1sini = 0.24$\,mas, which is much below the estimated Hipparcos precision, $2.66$\,mas. Even if $sini$ can be very small, the poor phase coverage of the Hipparcos data (30\,\%) makes it difficult to detect this signal. {To estimate the most probable orbital solution for this companion, we perform a Monte Carlo simulation (MCS). The MCS (G. Montagnier, PhD thesis; J. Hagelberg, private comm.) is very similar to the ones described in \citet{Wright-2008} and \citet{Brown-2004}. The idea is to generate all the possible Keplerian solutions and determine which ones most closely match the data. The method can be described in the following way :}
\begin{itemize}
\item We first calculate the $\chi^2$ of a polynomial fit to the data, $\chi^2_{poly}$. Here for HD7449, we use a second order polynomial.
\item We then choose a grid in the mass-period space to search for the solution.
\item For each point of this grid, which gives us already two parameters for the Keplerian model, we select randomly 1000 values of the four remaining parameters of the Keplerian ($T_0$, $\omega$, $\gamma$, and $e$). Each parameter is drawn from an uniform probability distribution except for the eccentricity which is drawn from the linear probability distribution $f(e) = 2e$ \citep{Duquennoy-1991}.
\item For each drawing, we retain the solution if $\chi^2_{Keplerian} - \chi^2_{poly} < 39.5$ ($5\sigma$ tolerance).
\item Each valid solution is then placed in a mass-period diagram where the color corresponds to the density of valid solution (dark = high density). The 1$\sigma$ and 2$\sigma$ contours incorporate 68\% and 95\% of valid solutions, respectively.
\end{itemize}
{The highest probability (see Fig. \ref{fig:5}) is found for a 4000-day period planet with a mass of $2\,M_J$, which closely corresponds to the properties of the fitted orbital solution. However, this solution is very poorly constrained when we consider the size of the 2$\sigma$ contour. Given the high eccentricity of the inner planet, it is very likely that a strong interaction occurs between the two companions, increasing the problem of dynamical stability {\citep[e.g][]{Fabrycky-2007}}. This system is therefore very interesting, and unfortunately much more data are needed to constrain it.}
%
\begin{figure}
\begin{center}
\includegraphics[width=8cm]{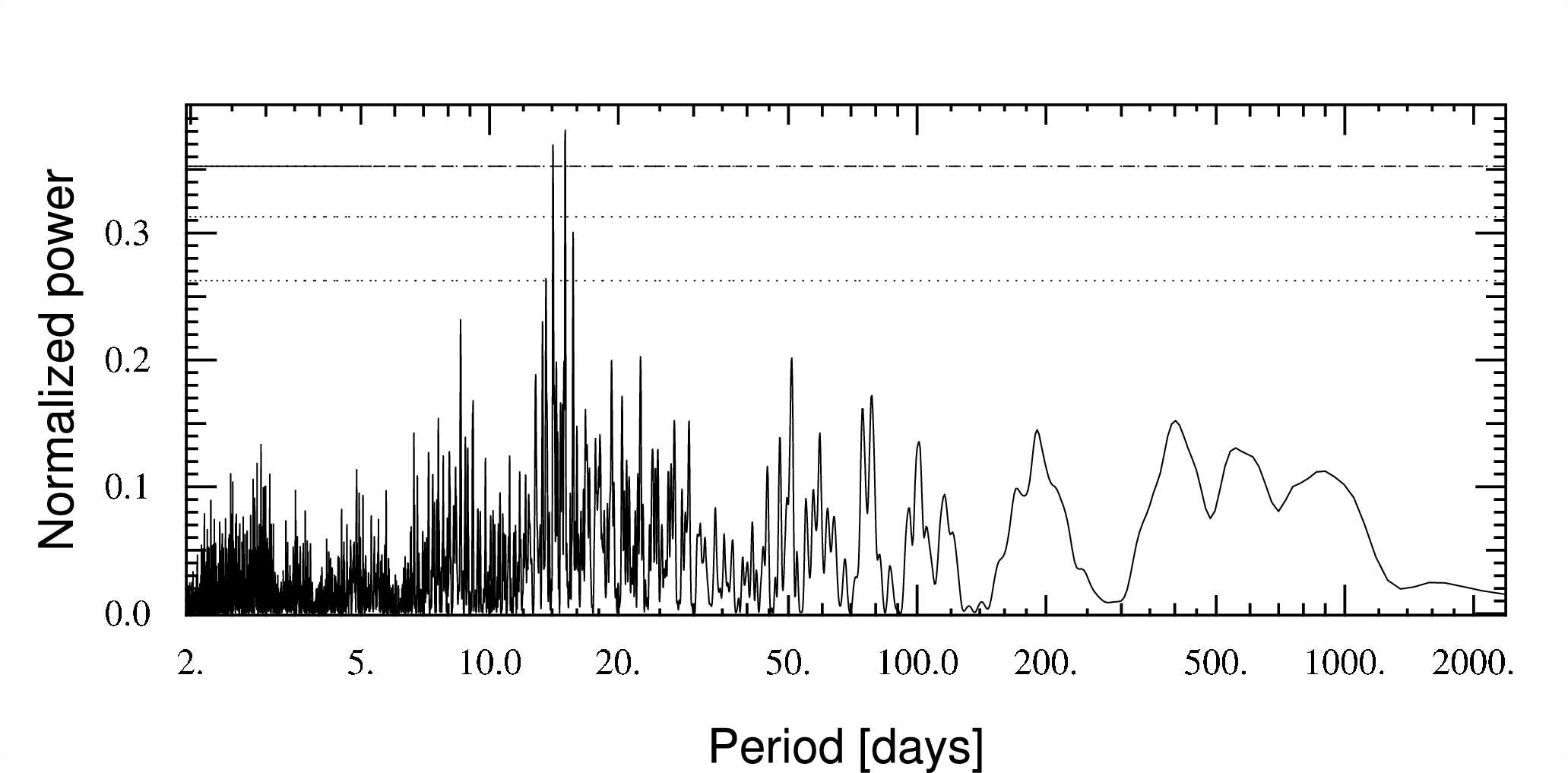}
\includegraphics[width=8cm]{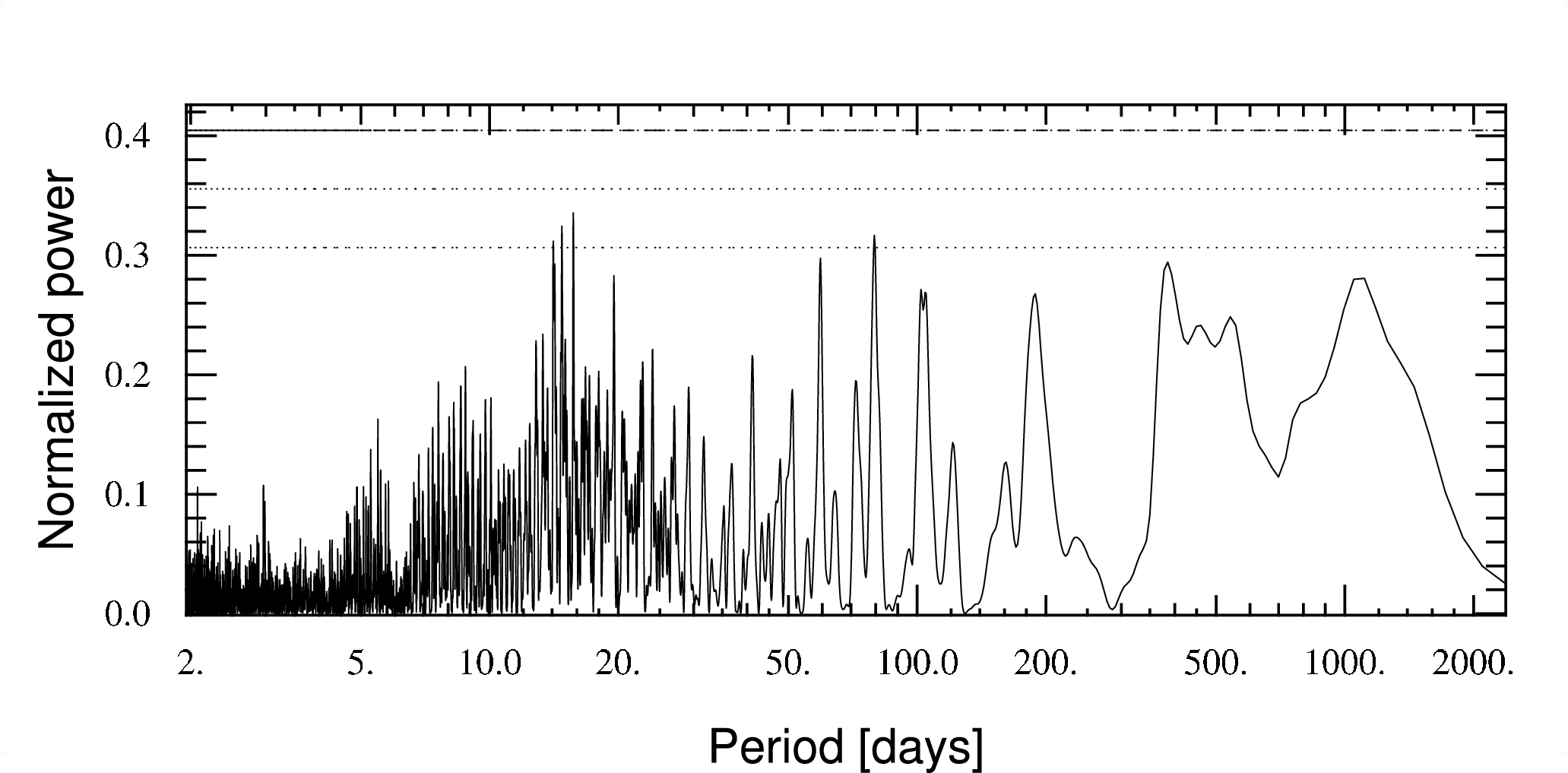}
\caption{All the periodograms are for HD7449. The horizontal lines correspond from top to bottom to a FAP of 0.1\,\%, 1\,\%, and 10\,\%. \emph{Top panel: }RV residuals after removing the two Keplerian orbital solution. \emph{Lower panel: }Activity index, log(R'$_{HK}$).}
\label{fig:4}
\end{center}
\end{figure}

\begin{figure*}
\begin{center}
\includegraphics[width=8cm]{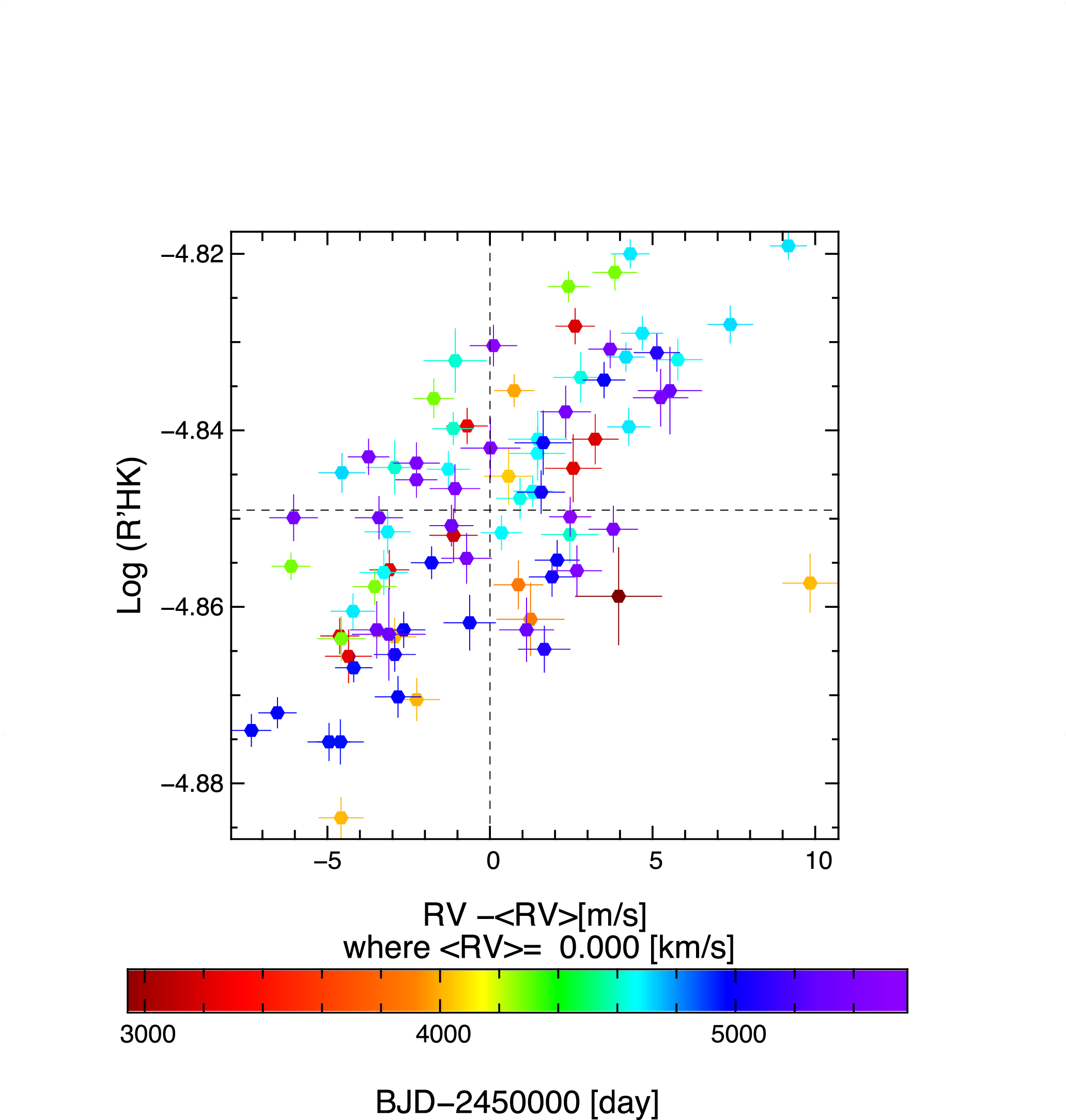}
\includegraphics[width=8cm]{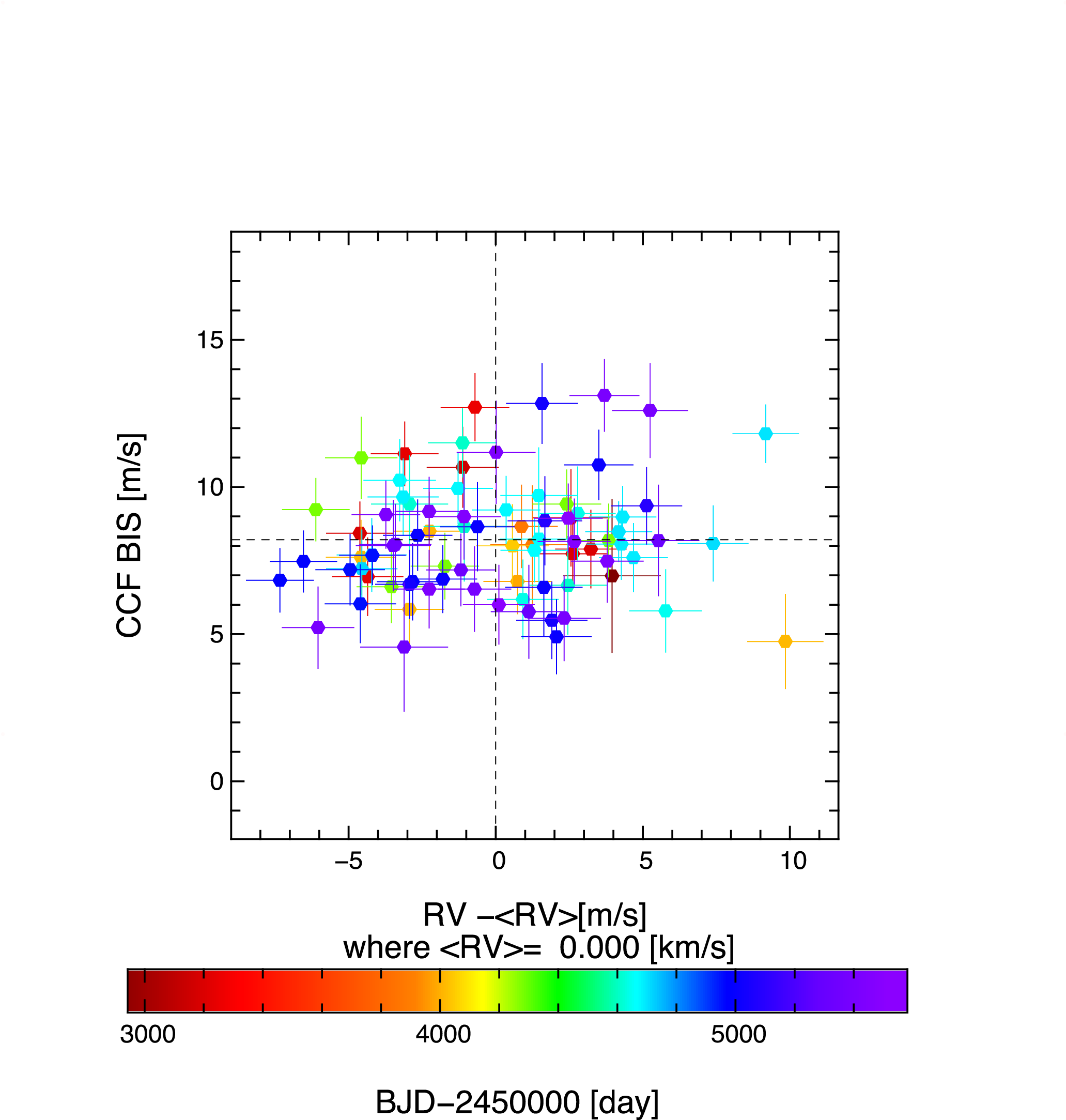}
\caption{Correlation for HD7449 between the activity index and the RV residuals (\emph{left}) and between the BIS span and the RV residuals (\emph{right}).}
\label{fig:6}
\end{center}
\end{figure*}

When looking at the periodograms of the RV residuals and the activity index (Fig. \ref{fig:4}), we found significant peaks around 14 days. These peaks are very significant in the RV periodogram, were we find a FAP smaller than 0.1\,\%, and weaker in the activity index periodogram. The estimated rotational period of the star, 13.30 $\pm$ 2.56 days (see Table \ref{tab:1}), closely matches this signature, which could therefore be associated with short-term activity. In addition, the activity origin of this signal is reinforced by the strong correlation between the RV residuals after removing the two planets, and the activity index (see Fig. \ref{fig:6} \emph{left}). {In the figure, measurements taken during the same time interval (same color) are spread throughout the correlation, which means that the activity has a short-term variation. Although we show for HD7199, HD137388, and HD204941 that a correlation is found between the RV residuals after removing the planets and the long-term variation of the activity index, this does not mean that a correlation between RV residuals and short-term activity variation is expected. Short-term activity is indeed induced by stellar rotation when magnetic regions are present on the star, whereas long-term activity is induced by the changing filling factor of magnetic regions during a magnetic cycle (see Sec. \ref{sect:1}). The two phenomena are therefore different and we cannot apply here the conclusion found for HD7199, HD137388, and HD204941}. As shown by several authors \citep[e.g][]{Boisse-2011,Queloz-2009,Huelamo-2008,Queloz-2001}, short-term activity can be highlighted by an anti-correlation between the RV residuals and the BIS span. Unfortunately, this anti-correlation is not seen in this case (see Fig. \ref{fig:6} \emph{right}).

%

{It is possible that the BIS span measurements are not sensitive enough to detect the RV-BIS span anti-correlation. However, without this anti-correlation, we cannot attest that activity is responsible for this short-term RV variation}. We therefore investigate the possibility of a short-period planet in the system. The orbital solution found for a three-companion system reduces the chi-square from 6.73 to 4.58 and the RV residual rms down to 2.94\,m\,s$^{-1}$ (only calculated on HARPS data), with a 14-day period planet. To confirm the existence of this planet, the induced signal should have the same phase for different observational periods. We therefore compare the circular orbital solution fitted to all the RV residuals and the ones fitted to different  RV residual chunks\footnote{RV residuals from the two Keplerian orbital solution shown in Table \ref{tab:2}.}. The circular orbit fit to the entire data set gives us the period of the signal, $P_{all\,data}$, the semi-amplitude, $K_{all\,data}$, and the time at periastron, $T_{0,all\,data}$. Using $P_{all\,data}$ for the period and a null eccentricity, we fit a Keplerian to each selected chunk and compare the values of $K$ and $T_0$ (seeTable \ref{tab:29}). If the signal is due to a planet, we should find for each chunk a $K$ and $T_0$ value compatible within the error bars with $K_{all\,data}$ and $T_{0,all\,data}$, respectively. The $T_0$ value for the second chunk is only compatible to a 3$\sigma$ level, therefore, given the strong RV-log(R'$_{HK}$) correlation, this 14-day period signal is more likely to be due to activity. We note that this technique is only valid for non-eccentric planets because the values of $T_0$ can be compared only in the case of circular orbits. We therefore cannot exclude an eccentric orbit for this 14-day period planet, although this is very unlikely for such a short-period.
\begin{table*}
\begin{center}
\caption{Time at periastron, $T_0$, and semi-amplitude for the 14.07 days signal present in the RV residuals of HD7449. Errorbars are Monte Carlo based 1$\sigma$ uncertainties.}  \label{tab:29}
\begin{tabular}{ccccc}
\hline
\hline Chunk & Nb of measurements & Estimated $T_0$ [days] & Semi-amplitude [m/s] \\
\hline
All data						& 82		& 2454719.6 $\pm$ 0.4		& 3.07  $\pm$ 0.49 \\
Chunk 1 (2454672 to 2454780) 	& 22		& 2454719.4 $\pm$ 0.6		& 4.03 $\pm$ 1.15 \\
Chunk 2 (2455020 to 2455162) 	& 17		& 2454722.0  $\pm$ 0.7		& 3.93 $\pm$ 1.18 \\
Chunk 3 (2455373 to 2455579)		& 20		& 2454718.9  $\pm$ 0.7		& 2.87 $\pm$ 0.91\\
\hline
\end{tabular} 
\end{center}
\end{table*}

In both cases, with or without removing the eventual 14-day period planet, the important residual rms indicates a significant short-term activity that can be explained by the short rotational period of the star.

\subsection{HD137388} \label{sect:5.3}

We collected 62 HARPS measurements for HD137388, between BJD =2453578 (July 26, 2005) and BJD = 2455632 (March 11, 2011). After removing three measurements with a SNR at 550 nm below 50, the mean SNR average at 550 nm is 90 (minimum is 51) and the mean RV uncertainty is 0.76 m\,s$^{-1}$ (photon and calibration noise).

\begin{figure*}
\begin{center}
\includegraphics[width=8cm]{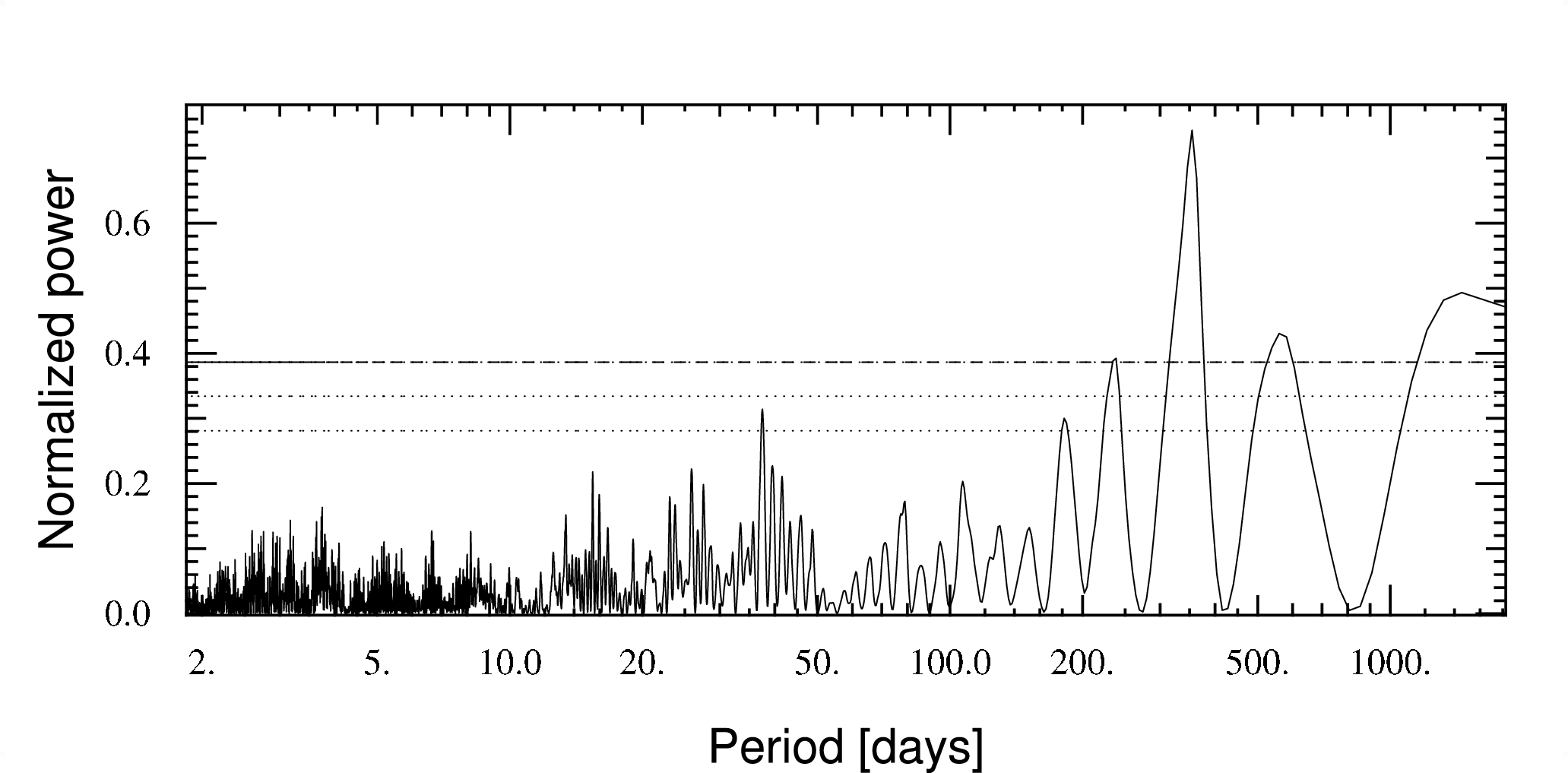}
\includegraphics[width=8cm]{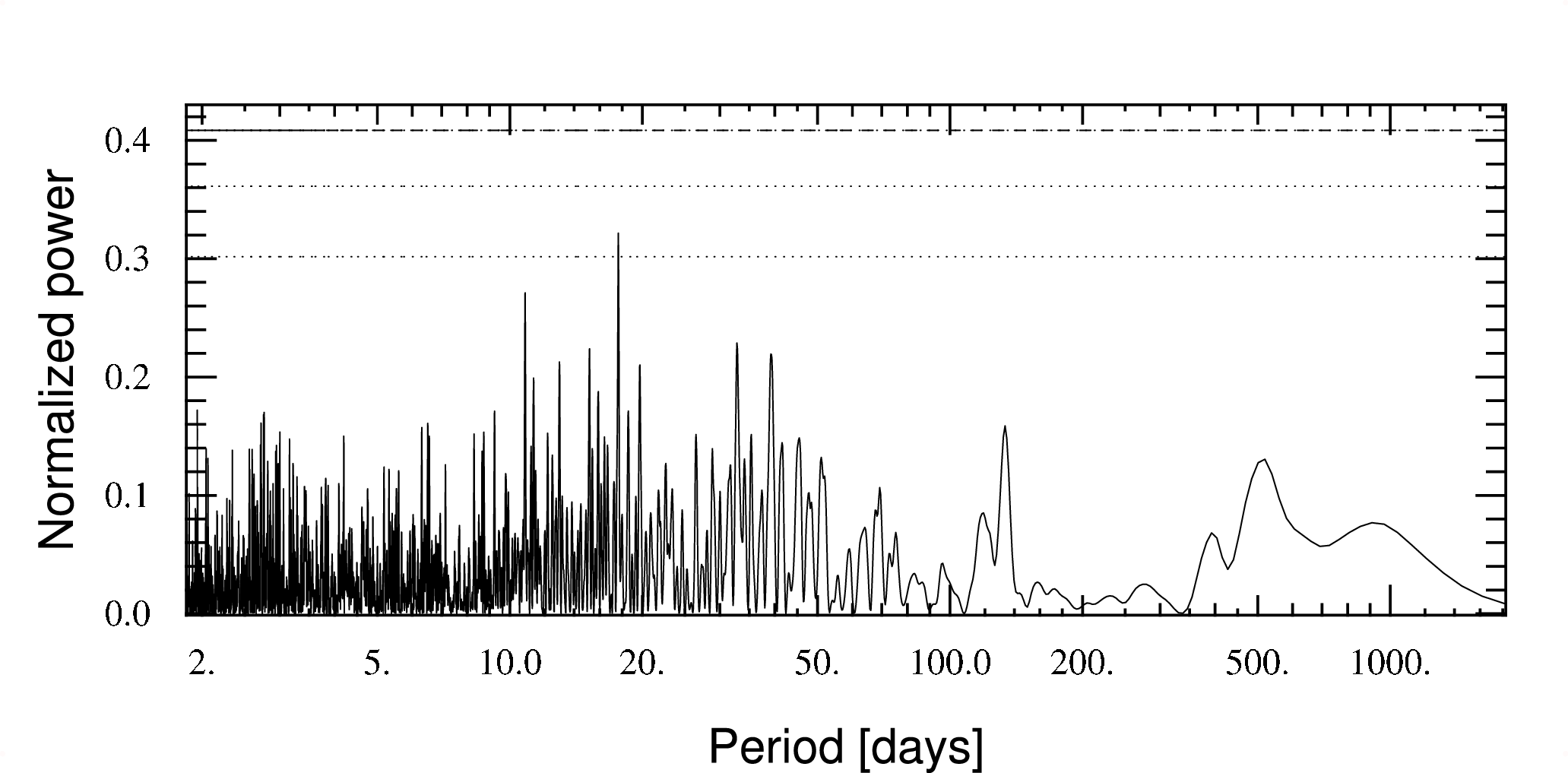}
\caption{Periodograms for HD137388 of the raw RVs (\emph{left}) and of the RV residuals after removing the planet and the effect of the magnetic cycle (\emph{right}). The horizontal lines correspond from top to bottom to a FAP of 0.1\,\%, 1\,\%, and 10\,\%.}
\label{fig:7}
\end{center}
\end{figure*}

\begin{figure*}
\begin{center}
\includegraphics[width=8cm]{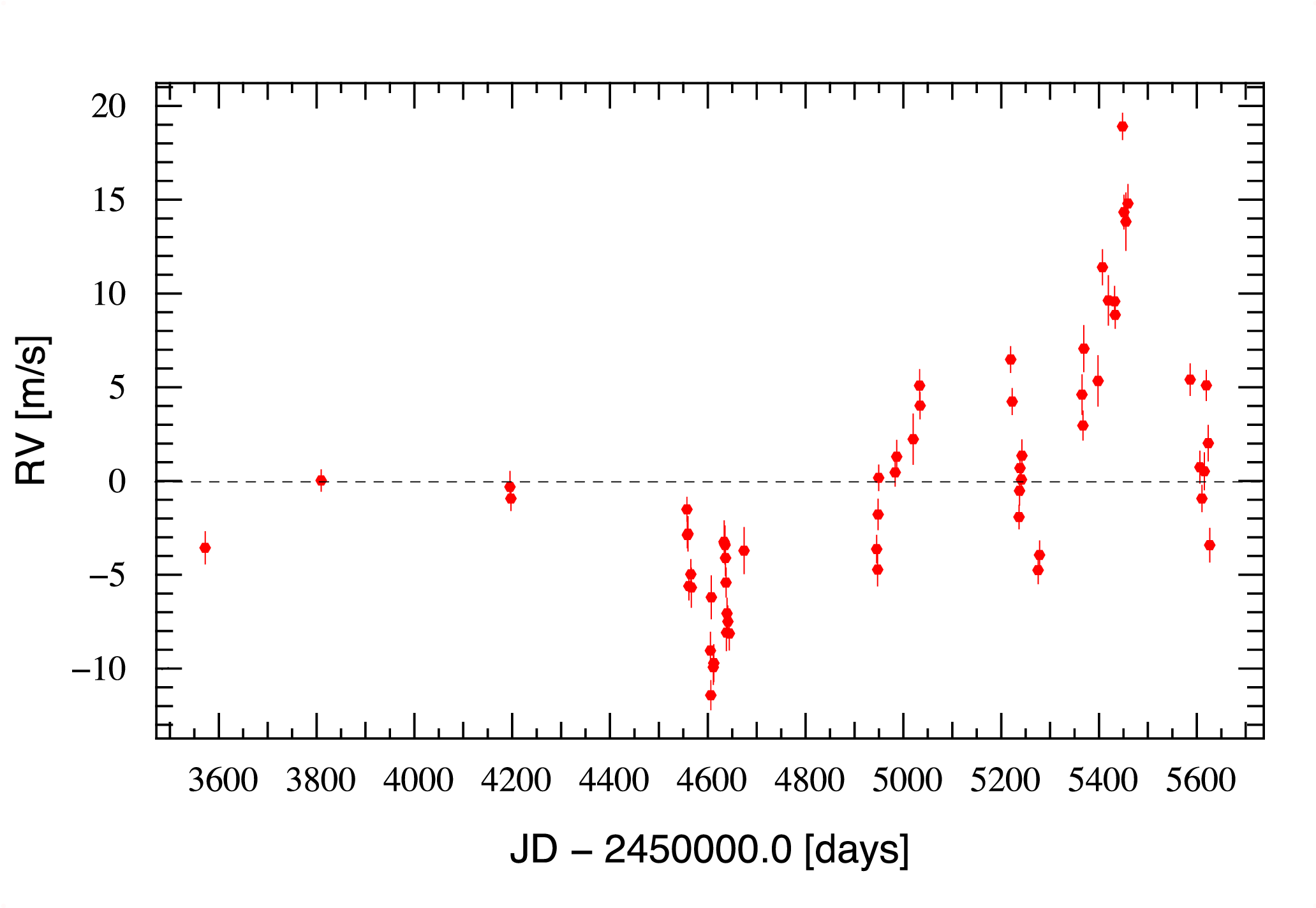}
\includegraphics[width=8cm]{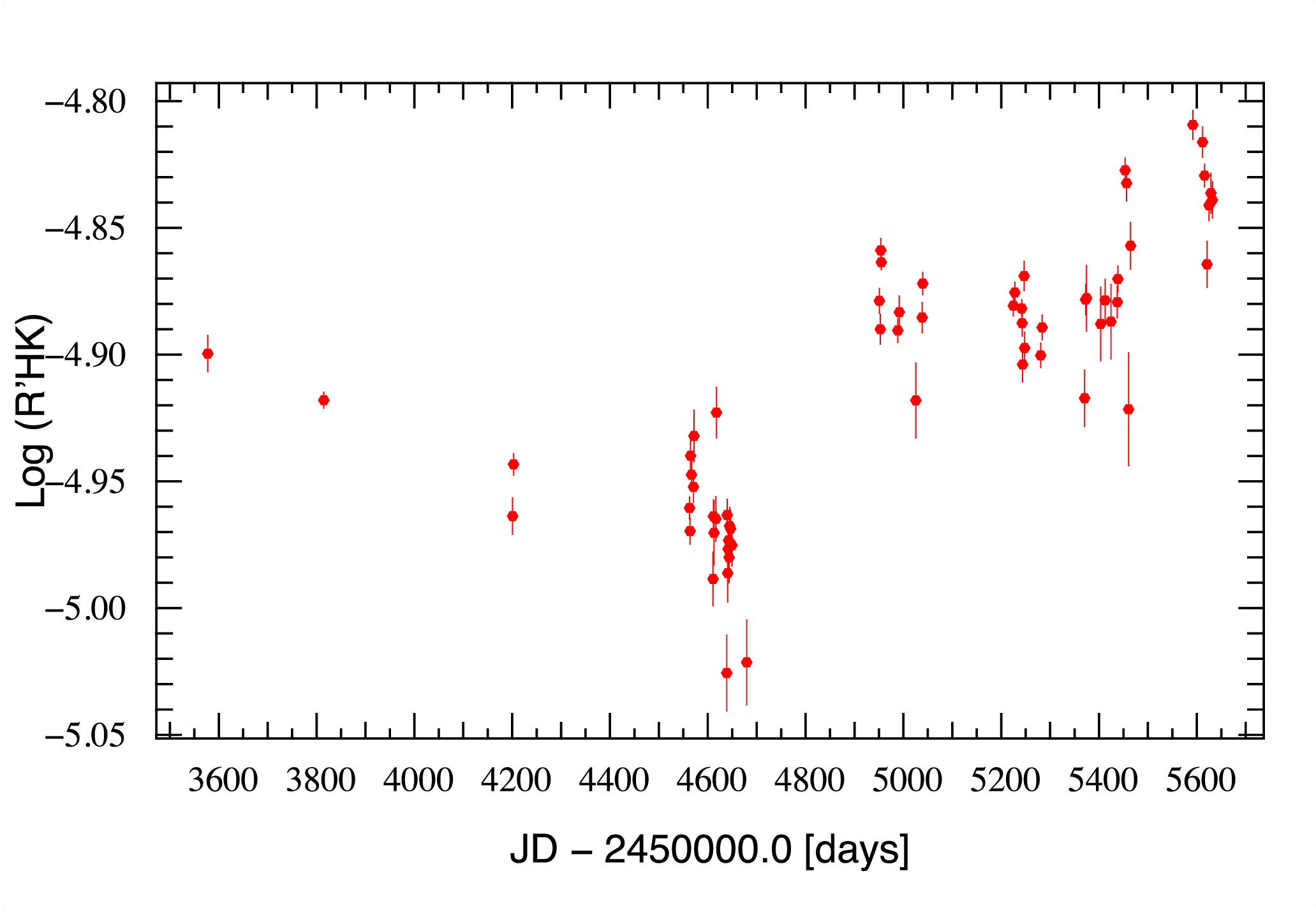}
\includegraphics[width=7.5cm]{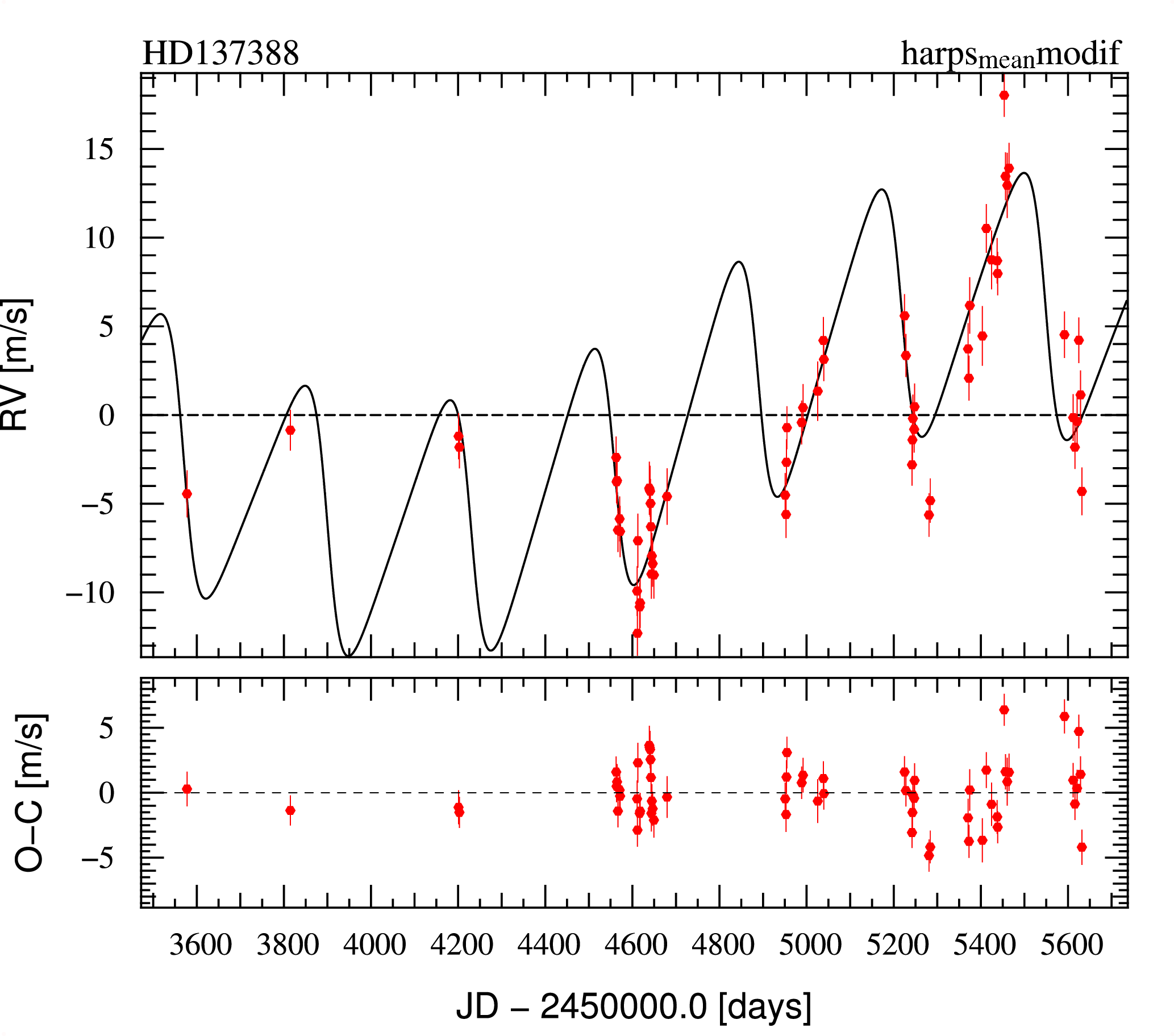}
\includegraphics[width=9cm]{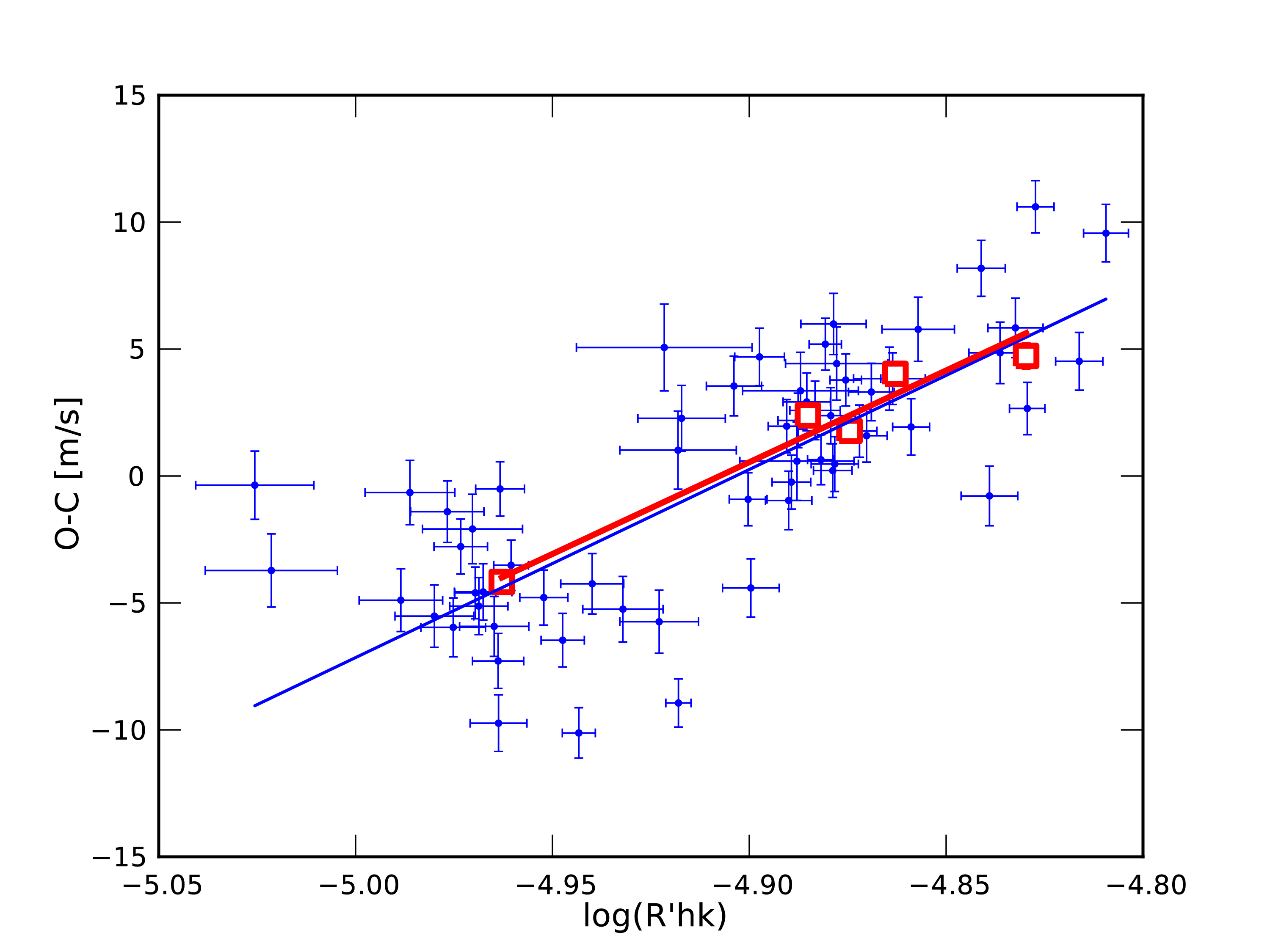}
\caption{All the graphs are for HD137388. \emph{Top panel: }RVs and the activity index, log(R'$_{HK}$), as a function of time. \emph{Lower left panel: }RV signal of the planet, in addition to the long-period variation associated to the magnetic cycle. \emph{Lower right panel: }Correlation between the RV residuals (O-C) after removing the planet and the activity index. Small points correspond to all the measurements, whereas large squares represent the same data binned over three months to average out short-term activity.}
\label{fig:8}
\end{center}
\end{figure*}

Studying the periodogram of the raw RVs (see left of Fig.\ref{fig:7}) we see a very significant peak at 346 days induced by the presence of a planet. In addition, some significant peaks are present at longer periods, which reflect a magnetic cycle that can clearly be identified from the activity index (see Fig. \ref{fig:8} \emph{top right}). The period of this cycle is difficult to estimate because we do not have a coverage of one complete period, although the period should be longer than 2000 days. {To determine the orbital solution for the planet taking into account the magnetic cycle, we apply the same strategy as for HD7199. We first fit a Keplerian model to the activity index (fixing the eccentricity to zero in this case) and then fit a two component Keplerian model to the RVs. One Keplerian fits the planet, with all the parameters free. The other one fits the magnetic cycle with all the parameters locked to the values found when fitting the activity index, except for the amplitude. The best-fit orbital solution infers a 330 day period planet with an eccentricity of 0.33 (see Fig. \ref{fig:8} \emph{lower left panel} and Table \ref{tab:2}).}

Looking at the periodogram of the RV residuals after removing the 330 day planet and the magnetic cycle (right of Fig.\ref{fig:7}), we do not find any peak with a FAP lower than 5\%, excluding the presence of another planet at this precision level.

As for HD7199, in Fig. \ref{fig:8} \emph{lower right panel}, we see a clear correlation between the RV residuals after removing the planet and the activity index, which confirms that an activity variation induced by a magnetic cycle influences the RV measurements.


\subsection{HD204941} \label{sect:5.4}

After six years of follow up, 35 spectra of HD204941 have been taken with HARPS. The measurements spanning from BJD = 2453342 (December 2, 2004) to BJD = 2455522 (November 21, 2010) have a mean SNR of 120 (minimum is 74). The mean photon and calibration noise uncertainty obtained for the RVs is 0.67 m\,s$^{-1}$. 

The raw RVs exhibit a huge excess of power near 1800 days in the periodogram induced by the presence of a Saturn-mass planet. {In addition, the activity index (Fig. \ref{fig:9} \emph{top right panel) varies on a long-period timescale. The measurements of HD204941 therefore display an incomplete solar-like magnetic cycle with a minimum amplitude of 0.1 dex and a minimum period between five and six years}. The planet signal is even more clear in the periodogram when subtracting from the RVs a quadratic drift similar to the activity index variation (Fig. \ref{fig:10} \emph{top}).
The best-fit orbital solution (see Fig. \ref{fig:9} \emph{middle panel} and Table \ref{tab:2}) corresponds to a 0.27\,$M_J$ mass planet with a period of 1733 days and an eccentricity of 0.37. {The properties of this planet is the least well constrained of this paper because the period is only constrained by a single epoch of measurements (the two first points). Thus we performed a Monte Carlo simulation (MSC) to estimate the most probable orbital solution (see Sect. \ref{sect:5.2}). Looking at the results of the MCS in Fig. \ref{fig:10}, we can see that the fitted solution is within the 1$\sigma$ contour and that it is well constrained according to the small size of the region delimited by the 1$\sigma$ contour.}

As for HD7199 and HD137388, we can see in Fig. \ref{fig:9} (\emph{lower right panel}) that a straightforward correlation is emphasized between the activity index and the RV residuals after removing the motion of the planet. Therefore magnetic cycles perturb RV measurements.
\begin{figure*}
\begin{center}
\includegraphics[width=8cm]{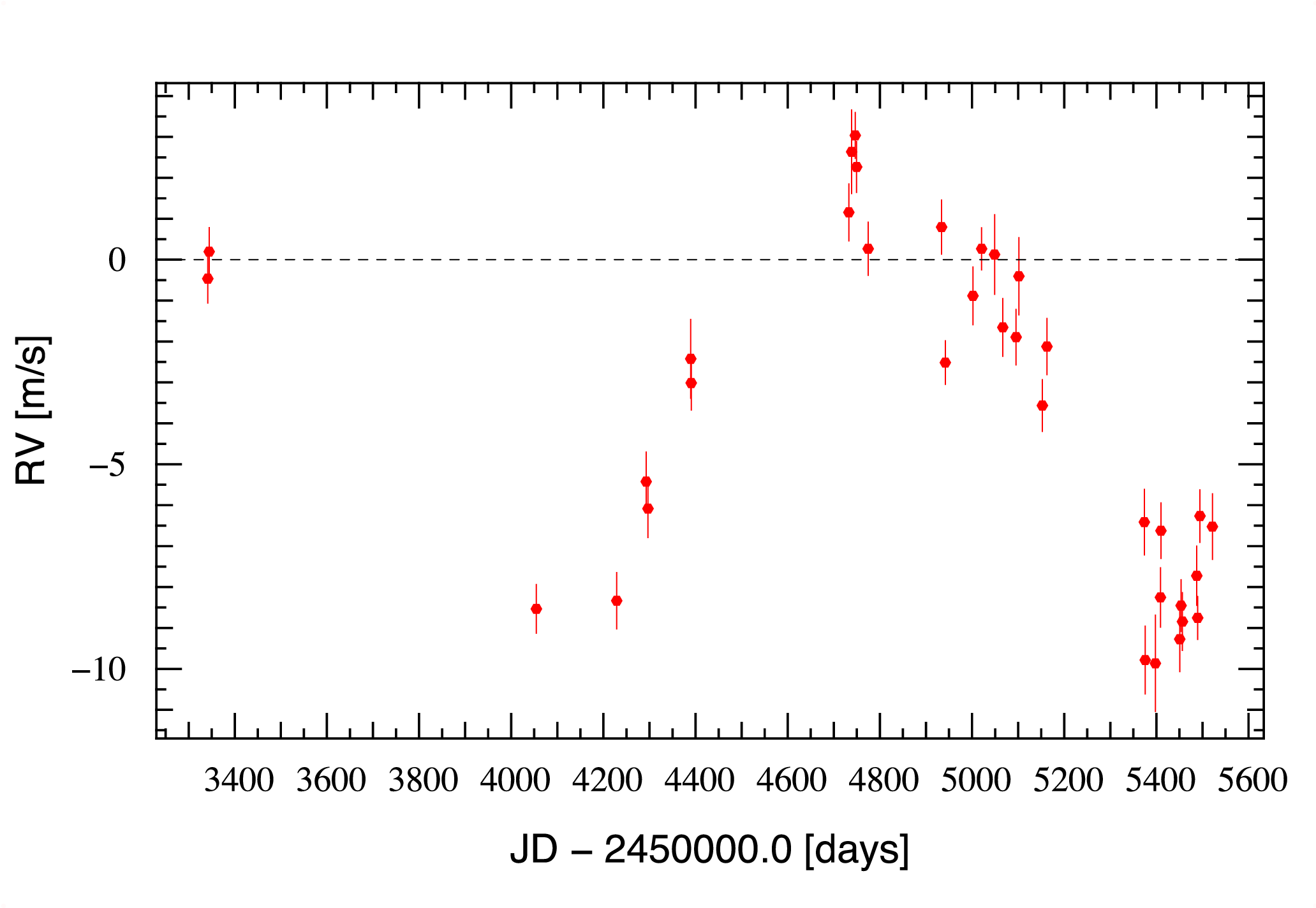}
\includegraphics[width=8cm]{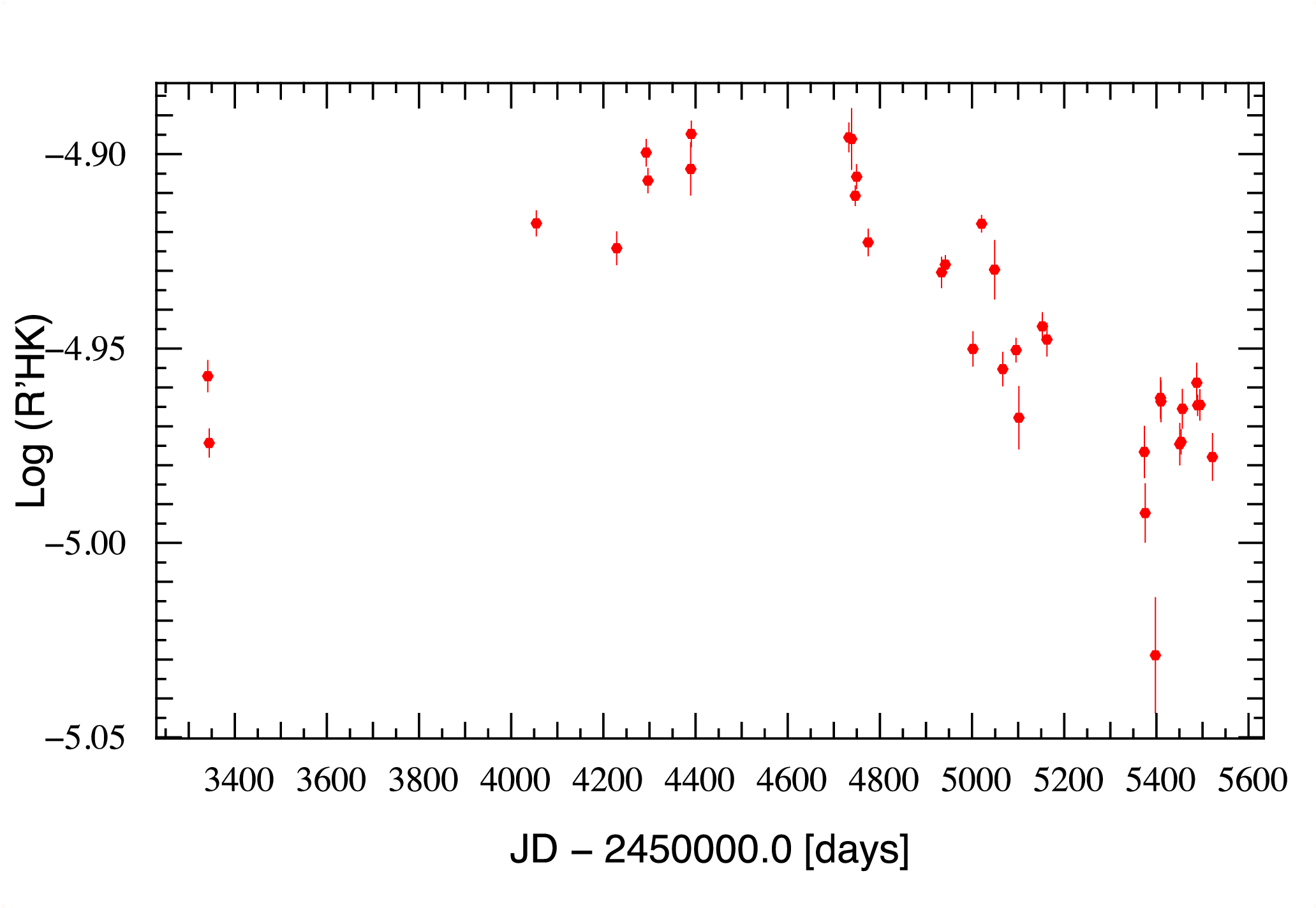}
\includegraphics[width=7.5cm]{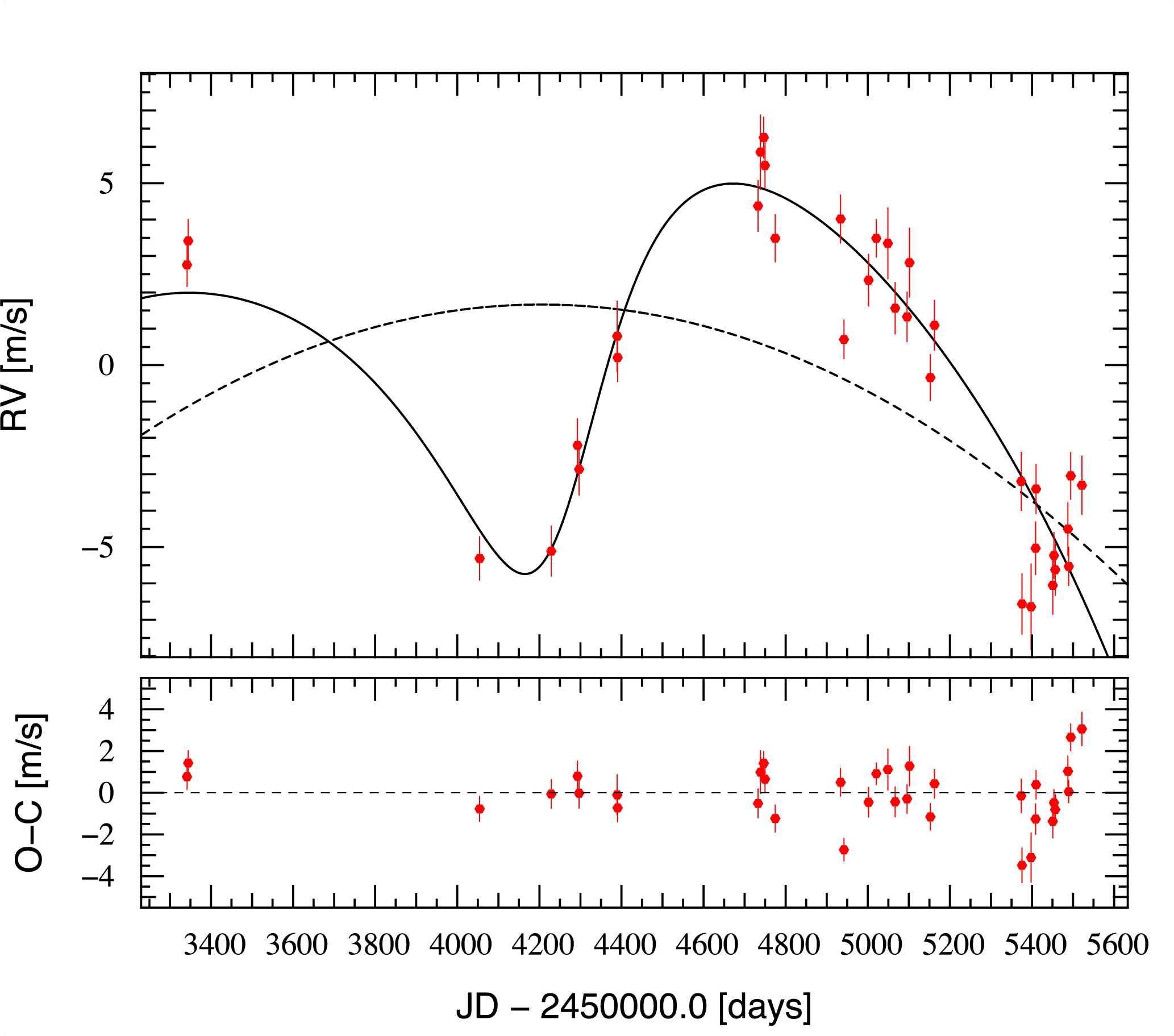}
\includegraphics[width=9cm]{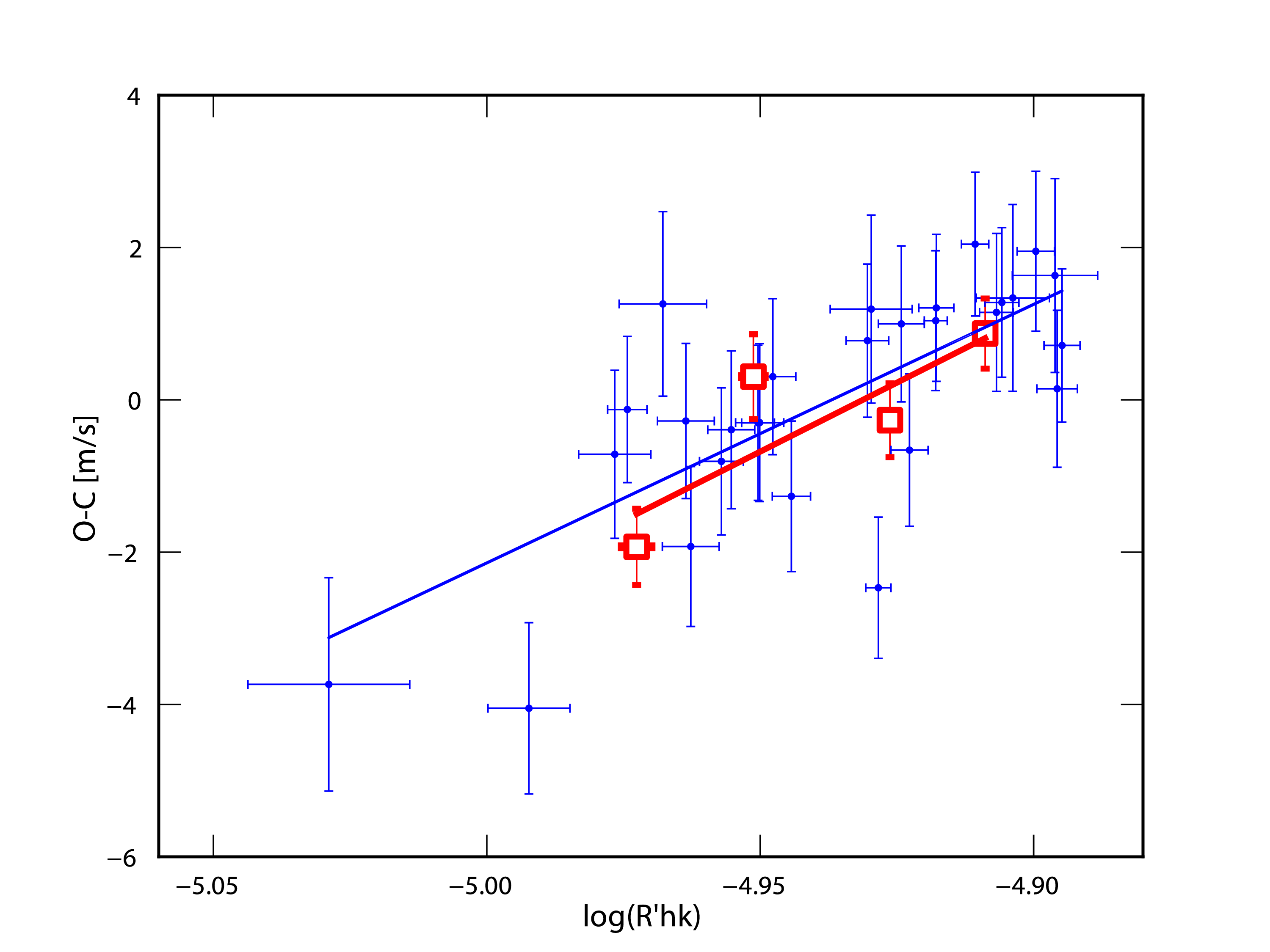}
\caption{All the graphs are for HD204941. \emph{Top panel: }RVs and the activity index, log(R'$_{HK}$), as a function of time. \emph{Lower left panel: }RV signal of the planet (continuous line) plus the quadratic drift fitting the magnetic cycle (dashed line). \emph{Lower right panel: }Correlation between the RV residuals (O-C) after removing the planet and the activity index. Small points correspond to all the measurements, whereas large squares represent the same data binned over three months to average out short-term activity.}
\label{fig:9}
\end{center}
\end{figure*}
\begin{figure*}
\begin{center}
\includegraphics[width=8cm]{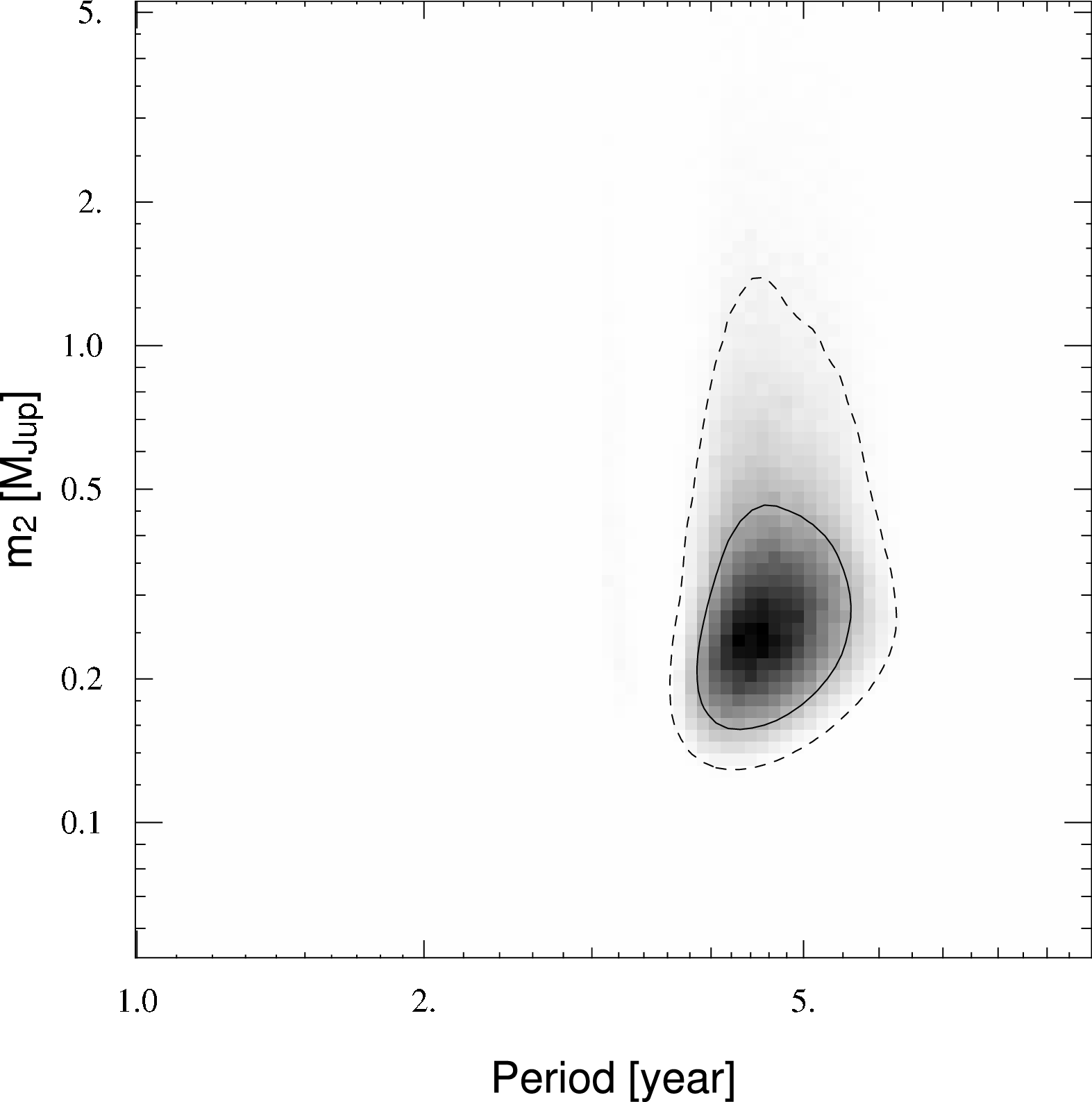}
\caption{Monte Carlo simulation of the companion of HD204941. The continuous contour corresponds to a level of 1\,$\sigma$ and the dashed one to 2\,$\sigma$ (see Sect. \ref{sect:5.2} for the definition of $\sigma$).}
\label{fig:10}
\end{center}
\end{figure*}
\begin{figure}
\begin{center}
\includegraphics[width=8cm]{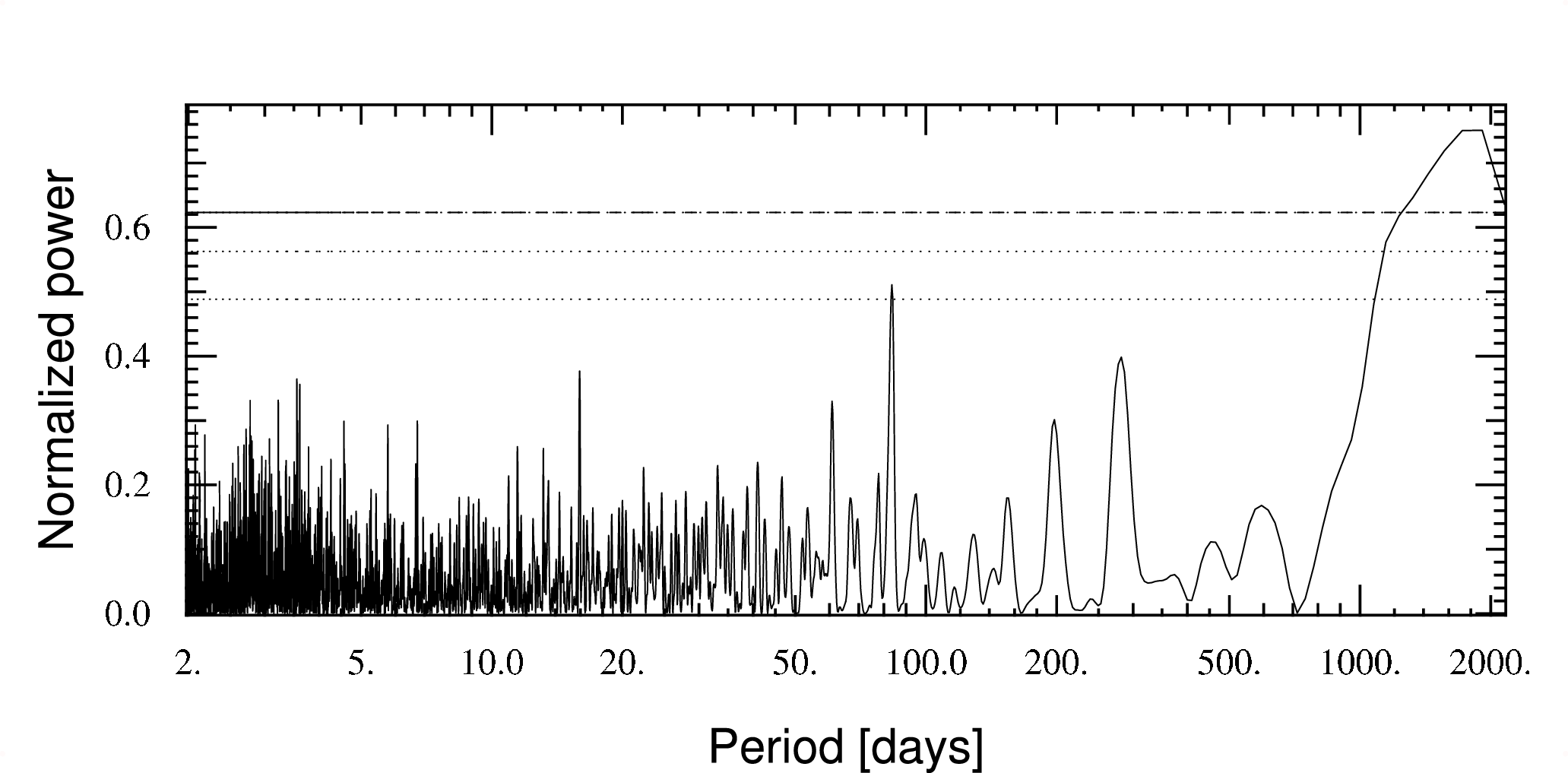}
\includegraphics[width=8cm]{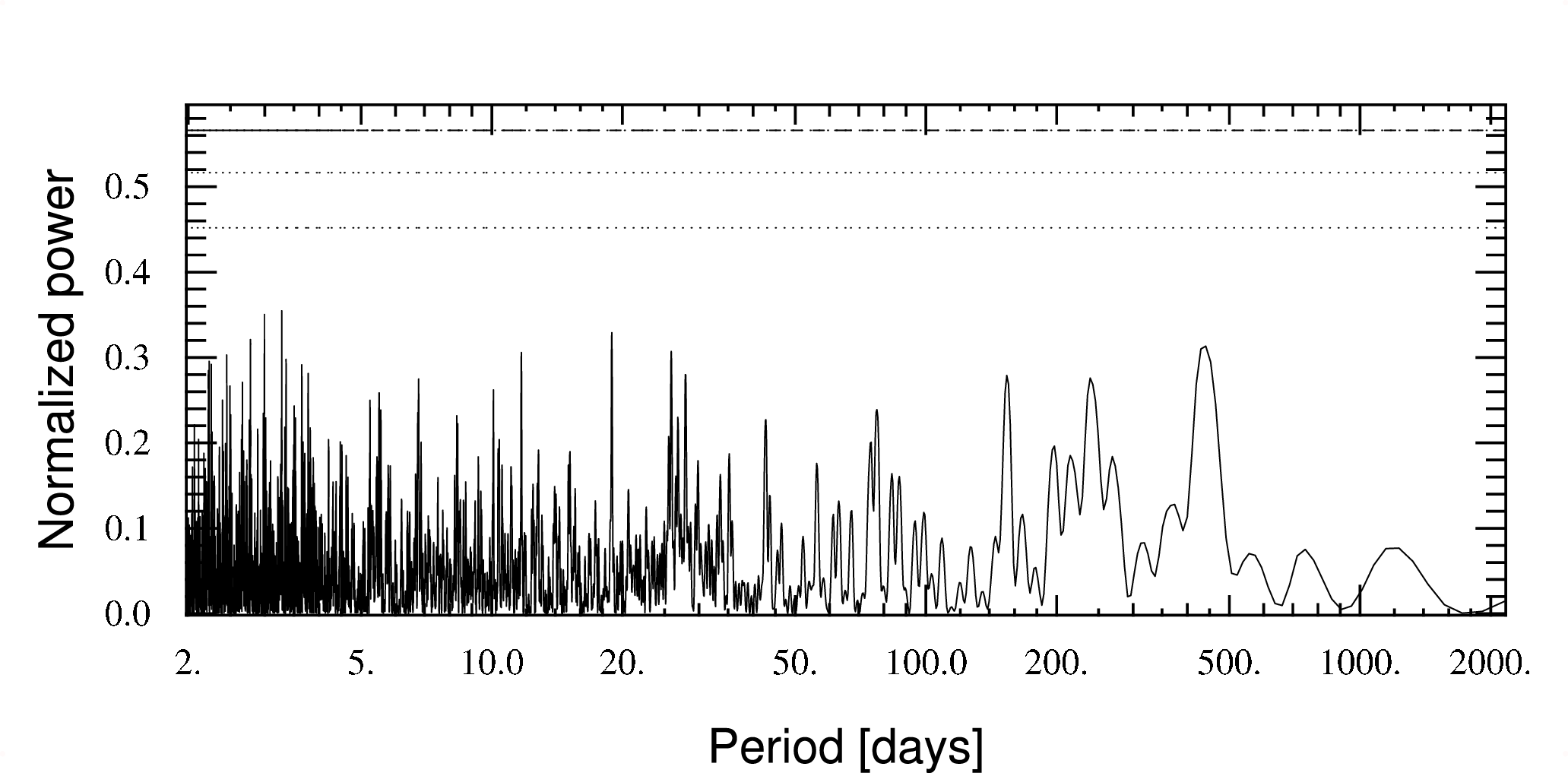}
\caption{Periodograms for HD204941 of the raw RVs after removing a quadratic drift (\emph{top}) and of the RV residuals after removing the planet and the quadratic drift (\emph{bottom}). The horizontal lines correspond from top to bottom to a FAP of 0.1\,\%, 1\,\%, and 10\,\%.}
\label{fig:11}
\end{center}
\end{figure}

If we consider the RV residual periodogram, Fig. \ref{fig:11} \emph{lower panel}, we do not find any peak with a FAP less than 50\%, excluding at this level of precision the presence of a second companion.

%
\begin{table*}
\begin{center}
\caption{Orbital solutions for HD7199b, HD7449b,c, and HD204941. Errorbars are Monte Carlo based 1$\sigma$ uncertainties
 and $\sigma$(O-C) m.s$^{-1}$ is the weighted rms  of the RV residuals.}  \label{tab:2}
\begin{tabular}{cccccc}
\hline
\hline Parameters & HD7199b & HD7449b & HD7449c & HD137388 & HD204941b \\
\hline
$P$ [days] 				& 615 $\pm$ 7											& 1275$^{+15}_{-11}$				& 4046$^{+186}_{-366}$					& 330$^{+2}_{-4}$										& 1733 $\pm$ 74\\
$T$ [JD-2400000] 			& 55030 $\pm$ 277										& 55298$^{+31}_{-22}$				& 55883$^{+85}_{-172}$					& 55209$^{+23}_{-28}$									& 56015 $\pm$ 92\\
$e$ 						& 0.19$^{+0.19}_{-0.13}$									& 0.82 $\pm$ 0.06					& 0.53 $\pm$ 0.08						& 0.36 $\pm$ 0.12										& 0.37 $\pm$ 0.08\\
$\omega$ [deg] 			& 155$^{+90}_{-64}$									& -21 $\pm$ 6						& 11 $\pm$ 8							& 86$^{+23}_{-35}$										& -132 $\pm$ 18\\
$K$ [m\,s$^{-1}$] 			& 7.76  $\pm$ 0.58										& 41.59$^{+10.79}_{-19.52}$			& 30.04$^{+5.41}_{-7.21}$				& 7.94$^{+0.70}_{-1.05}$									& 5.94 $\pm$ 0.71\\
$V$ [km.s$^{-1}$] 			& 5.701  $\pm$ 0.001									& -19.705 $\pm$ 0.018				& -19.716 $\pm$ 0.006					& 26.351  $\pm$ 0.001									& 32.637 $\pm$ 0.001\\
$m$\,sin\,$i$\,[$M_{Jup}$]		& 0.290 $\pm$ 0.023									& 1.11$^{+0.27}_{-0.37}$				& 2.00$^{+0.36}_{-0.54}$					& 0.223 $\pm$ 0.029									& 0.266 $\pm$ 0.032\\
$m$\,sin\,$i$\,[$M_{Earth}$]	& 92 $\pm$ 7											& 353$^{+86}_{-118}$				& 636$^{+114}_{-172}$					& 71 $\pm$ 9											& 85 $\pm$ 10\\
$a$ [AU] 					& 1.36 $\pm$ 0.02										& 2.30 $\pm$ 0.04					& 4.96$^{+0.20}_{-0.30}$					& 0.89 $\pm$ 0.02										& 2.56 $\pm$ 0.08\\
\hline						
$N_{meas}$ 				& 88													& \multicolumn{2}{c}{142}													& 62													&35\\
Span [days] 				&  2579												&  \multicolumn{2}{c}{4061}												& 2054												&2180\\
$\sigma$(O-C) [m.s$^{-1}$] 	& 2.63												& \multicolumn{2}{c}{3.81 (HARPS only)}										& 2.39												& 1.31\\
$\chi^2_{red}$ 				& 5.23 $\pm$ 0.36										& \multicolumn{2}{c}{6.73 $\pm$ 0.31}										& 3.76 $\pm$ 0.37										& 1.61 $\pm$ 0.35\\
\hline		
\end{tabular}
\end{center}
\end{table*}

\section{Discussion and conclusion} \label{sect:6}

We have presented the detections and the orbital parameters of four planets and a companion of unknown type in four different systems. Three of these planets, HD137388 (67 M$_{\oplus}$, 331 days), HD204941b (85 M$_{\oplus}$, 1733 days), and HD7199b (92 M$_{\oplus}$, 615 days) have low masses and long orbital periods in comparison with known planets\footnote{See The Extrasolar Planets Encyclopaedia, http://exoplanet.eu}. Only three other planets
have an orbital period longer than 300 days and a mass lower than 100 M$_{\oplus}$ : HD10180\,g \citep[21 M$_{\oplus}$, 601 days,][]{Lovis-2010} , HD85390\,b \citep[42 M$_{\oplus}$, 788 days,][]{Mordasini-2011}, and HD10180\,h (64 M$_{\oplus}$, 2222 days). We note that all these planets have been found using HARPS, thus this instrument is exploring a new domain of sub-Saturn planets beyond $\sim$\,1\,AU.
The fourth planet, HD7449b, has a high eccentricity of 0.82 and complements the five already known planets with a higher eccentricity. Although the eccentricity is very high, the transiting probability is low because of the large separation between the star and the periastron of the planet orbit ($\sim$\,0.6\,\% for the transit and 1\,\% for the anti-transit\footnote{The angle of the periastron, $\omega$, has been taken into account in the probability computation \citep[][]{Kane-2008}}). The discovery of these four new planets is very important to constrain and test models of planetary formation because they occupy regions of the mass-period-eccentricity space where only a few other discoveries have been made until now.

We found a second companion orbiting HD7449. Given that the data only cover a small fraction of the orbital period, the nature of this object, planet, brown dwarf, or small stellar companion is hard to determine. A Monte Carlo simulation shows, however, that a planetary companion with a mass similar to 2\,$M_J$ and a period close to 4000 days is the most probable. {Given the high eccentricity of the inner planet, it is very likely that a strong interaction occurs between the two companions, although much more data are needed to define properly this interesting multi-planetery system.}


The planets announced in this paper for the first time have been discovered even though their host stars display clear signs of activity. We have found that HD7449 exhibits signs of short-term activity, whereas HD7199, HD137388, and HD204941 have solar-like magnetic cycles.

When we examined the periodogram of HD7449 RV residuals we are able to identify two significant peaks close to a 14 day period, which is the estimated rotational period of the star, 13.30 $\pm$ 2.56 days. A similar signal appears in the periodogram of the activity index, although does not appear to be very significant. In addition, no signal was detected in either the BIS span or the FWHM. Therefore, a low-mass planet might be present. Looking at different chunks to see whether the 14-day period signal fitted stays in phase or not, we have arrived at the conclusion that the main signal is most probably caused by short-term activity and not induced by a planet on a circular orbit. However, a low-mass eccentric planet might still be present at this period and more measurements are needed to clearly define the nature of this 14-day signal.

When examining the RVs and the fitted planets for HD7199, HD137388, and HD204941, it is clear that magnetic cycles induce RV variations that could be misinterpreted as long-period planetary signature. Therefore, the long-term variations in the activity index have to be studied properly to distinguish between the real signature of a planet and long-term activity noise. When the long-term activity index variation evolves smoothly, it can be readily fitted and thus effect of magnetic cycles can be removed from RVs. This has been done for HD7199, HD137388, and HD204941 and gives a good estimation of the planet parameters, with rather small errors. 

{In \citet{Meunier-2010a}, the authors argued that the Sun should show RV variations of 10\,m\,s$^{-1}$ over its cycle. In this paper, we have confirmed that this behavior is seen on other solar-like stars and one can ask whether all stars showing magnetic cycles have a long-term RV variation induced by the long-term activity variation. The high precision HARPS sample, composed of 451 stars, provides a good set of measurements to search for this activity-RV correlation. However, owing to irregular sampling, this property can only be studied for a small set of stars. In preliminary work \citep[][]{Dumusque-2010}, we demonstrated that the magnetic cycle-RV correlation can be observed for 31 comprehensively observed stars and that in addition, RVs of G dwarfs are more sensitive to magnetic cycles than RVs of K dwarfs. A more complete study is in progress and will be soon published (Lovis et al. in prep.).}


\begin{acknowledgements}

We are grateful to all technical and scientific collaborators of the HARPS Consortium, ESO Headquarters and ESO La Silla who have contributed with their extraordinary passion and valuable work to the success of the HARPS project. We would like to thank the Swiss National Science Foundation for its continuous support. XD and NCS acknowledge the support by the European Research Council/European Community under the FP7 through Starting Grant agreement number 239953. NCS also acknowledges the support from Funda\c{c}\~ao para a Ci\^encia e a Tecnologia (FCT) through programme Ci\^encia\,2007 funded by FCT/MCTES (Portugal) and POPH/FSE (EC), and in the form of grants reference PTDC/CTE-AST/098528/2008 and PTDC/CTE-AST/098604/2008. XD would like to thanks J. Hagelberg, J. Sahlmann and A. H. M. J. Triaud for useful work and comments on the paper.

\end{acknowledgements}

\onllongtab{4}{
\begin{longtable}{ccccc}
\caption{\label{test} RVs and activity index for HD7199.}\\ 
\hline
\hline
Julian date  & RV & RV error & log(R'$_{HK}$) & log(R'$_{HK}$) error\\ 
$[\mathrm{T} - 2400000]$ & $[\mathrm{km}\,\mathrm{s}^{-1}]$ & $[\mathrm{km}\,\mathrm{s}^{-1}]$ & - & -\\ 
\hline
\endfirsthead
\caption{Continued.} \\ 
\hline
Julian date  & RV & RV error & log(R'$_{HK}$) & log(R'$_{HK}$) error\\ 
$[\mathrm{T} - 2400000]$ & $[\mathrm{km}\,\mathrm{s}^{-1}]$ & $[\mathrm{km}\,\mathrm{s}^{-1}]$ & - & -\\ 
\hline
\endhead
\hline
\endfoot
\hline
\endlastfoot
        52945.64331 & 5.68668 & 0.00158 & -5.03620 & 0.02460 \\ 
        53000.62100 & 5.68655 & 0.00115 & -5.02860 & 0.00580 \\ 
        53287.70413 & 5.70180 & 0.00066 & -4.99570 & 0.00590 \\ 
        53289.71645 & 5.70252 & 0.00047 & -4.98510 & 0.00310 \\ 
        53294.70385 & 5.70341 & 0.00069 & -5.03540 & 0.00660 \\ 
        53307.65345 & 5.70406 & 0.00055 & -5.01930 & 0.00450 \\ 
        53573.92818 & 5.69920 & 0.00062 & -4.89130 & 0.00440 \\ 
        53575.88374 & 5.69604 & 0.00051 & -4.89510 & 0.00310 \\ 
        53668.65336 & 5.69917 & 0.00050 & -4.87460 & 0.00280 \\ 
        53672.71856 & 5.70056 & 0.00043 & -4.90130 & 0.00240 \\ 
        53675.72270 & 5.70174 & 0.00044 & -4.89440 & 0.00230 \\ 
        53694.59052 & 5.70364 & 0.00043 & -4.86760 & 0.00280 \\ 
        53722.58521 & 5.71868 & 0.00045 & -4.81300 & 0.00210 \\ 
        53759.52940 & 5.71064 & 0.00045 & -4.81810 & 0.00220 \\ 
        53763.53228 & 5.70690 & 0.00040 & -4.85500 & 0.00190 \\ 
        53944.90474 & 5.71087 & 0.00064 & -4.89060 & 0.00330 \\ 
        53948.91953 & 5.71293 & 0.00087 & -4.86840 & 0.00610 \\ 
        53950.91116 & 5.72088 & 0.00047 & -4.85220 & 0.00240 \\ 
        53974.80652 & 5.71049 & 0.00064 & -4.89120 & 0.00330 \\ 
        53982.84136 & 5.71237 & 0.00053 & -4.86720 & 0.00210 \\ 
        54047.58122 & 5.71398 & 0.00048 & -4.83410 & 0.00230 \\ 
        54049.59909 & 5.71392 & 0.00054 & -4.85110 & 0.00290 \\ 
        54051.62771 & 5.71048 & 0.00061 & -4.85040 & 0.00350 \\ 
        54053.67721 & 5.70635 & 0.00051 & -4.87130 & 0.00270 \\ 
        54076.67169 & 5.70067 & 0.00055 & -4.91310 & 0.00350 \\ 
        54077.60707 & 5.70378 & 0.00043 & -4.90090 & 0.00210 \\ 
        54079.61371 & 5.70509 & 0.00052 & -4.89350 & 0.00300 \\ 
        54083.61658 & 5.70763 & 0.00061 & -4.87830 & 0.00390 \\ 
        54121.53242 & 5.70344 & 0.00064 & -4.89550 & 0.00430 \\ 
        54340.77779 & 5.70648 & 0.00047 & -4.90150 & 0.00230 \\ 
        54341.84528 & 5.70376 & 0.00057 & -4.91310 & 0.00350 \\ 
        54342.72951 & 5.70496 & 0.00056 & -4.91420 & 0.00320 \\ 
        54674.92474 & 5.69992 & 0.00069 & -4.96750 & 0.00540 \\ 
        54681.87257 & 5.70383 & 0.00056 & -4.94280 & 0.00350 \\ 
        54683.90807 & 5.70574 & 0.00057 & -4.93270 & 0.00370 \\ 
        54702.83812 & 5.69882 & 0.00061 & -4.93110 & 0.00390 \\ 
        54705.83622 & 5.69810 & 0.00071 & -4.94660 & 0.00530 \\ 
        54708.84558 & 5.69509 & 0.00047 & -4.97060 & 0.00270 \\ 
        54736.77784 & 5.69529 & 0.00063 & -4.94540 & 0.00440 \\ 
        55037.94480 & 5.70292 & 0.00048 & -5.00810 & 0.00300 \\ 
        55038.85578 & 5.70278 & 0.00047 & -5.00830 & 0.00280 \\ 
        55039.90292 & 5.70290 & 0.00131 & -5.00480 & 0.01630 \\ 
        55064.83279 & 5.70396 & 0.00046 & -5.01810 & 0.00270 \\ 
        55066.84320 & 5.70550 & 0.00059 & -5.01200 & 0.00320 \\ 
        55070.79684 & 5.70354 & 0.00067 & -5.02840 & 0.00570 \\ 
        55074.79574 & 5.70436 & 0.00048 & -5.01440 & 0.00300 \\ 
        55077.76279 & 5.70546 & 0.00049 & -5.00640 & 0.00310 \\ 
        55095.71254 & 5.70411 & 0.00058 & -5.00160 & 0.00300 \\ 
        55098.67792 & 5.70853 & 0.00046 & -4.99770 & 0.00260 \\ 
        55100.70527 & 5.70604 & 0.00044 & -4.99480 & 0.00240 \\ 
        55103.74133 & 5.70178 & 0.00083 & -5.00680 & 0.00760 \\ 
        55372.91879 & 5.68687 & 0.00052 & -5.05170 & 0.00430 \\ 
        55374.92597 & 5.68942 & 0.00087 & -5.09250 & 0.01100 \\ 
        55375.89051 & 5.68724 & 0.00071 & -5.06460 & 0.00800 \\ 
        55396.88904 & 5.68357 & 0.00125 & -5.12170 & 0.02140 \\ 
        55401.82522 & 5.68802 & 0.00100 & -5.12330 & 0.01430 \\ 
        55403.83006 & 5.68449 & 0.00075 & -5.05840 & 0.00830 \\ 
        55408.92365 & 5.68995 & 0.00064 & -5.06500 & 0.00690 \\ 
        55410.85733 & 5.68690 & 0.00071 & -5.06730 & 0.00730 \\ 
        55414.83224 & 5.68638 & 0.00067 & -5.08440 & 0.00740 \\ 
        55423.79437 & 5.69207 & 0.00074 & -5.05820 & 0.00780 \\ 
        55425.72574 & 5.68976 & 0.00074 & -5.05820 & 0.00790 \\ 
        55426.89617 & 5.68673 & 0.00050 & -5.04830 & 0.00440 \\ 
        55427.83151 & 5.68827 & 0.00056 & -5.05760 & 0.00500 \\ 
        55428.80997 & 5.68603 & 0.00069 & -5.05570 & 0.00690 \\ 
        55435.77465 & 5.68665 & 0.00047 & -5.05300 & 0.00360 \\ 
        55436.79678 & 5.68718 & 0.00042 & -5.04790 & 0.00290 \\ 
        55437.70673 & 5.68872 & 0.00048 & -5.04580 & 0.00370 \\ 
        55439.83300 & 5.68872 & 0.00052 & -5.05410 & 0.00430 \\ 
        55443.75192 & 5.68904 & 0.00066 & -5.06550 & 0.00650 \\ 
        55444.75125 & 5.68806 & 0.00050 & -5.06700 & 0.00410 \\ 
        55446.81032 & 5.68706 & 0.00063 & -5.07110 & 0.00490 \\ 
        55450.83568 & 5.68752 & 0.00074 & -5.06910 & 0.00670 \\ 
        55454.75159 & 5.69135 & 0.00063 & -5.05030 & 0.00440 \\ 
        55455.73443 & 5.68958 & 0.00064 & -5.07070 & 0.00630 \\ 
        55463.72355 & 5.68718 & 0.00055 & -5.06910 & 0.00480 \\ 
        55480.68798 & 5.68745 & 0.00054 & -5.06290 & 0.00470 \\ 
        55482.73161 & 5.68909 & 0.00062 & -5.08130 & 0.00610 \\ 
        55483.62245 & 5.68844 & 0.00065 & -5.06250 & 0.00640 \\ 
        55485.66053 & 5.68839 & 0.00050 & -5.06760 & 0.00390 \\ 
        55492.69542 & 5.68871 & 0.00047 & -5.05490 & 0.00360 \\ 
        55493.68688 & 5.68812 & 0.00049 & -5.05510 & 0.00400 \\ 
        55495.72838 & 5.68771 & 0.00050 & -5.04680 & 0.00390 \\ 
        55497.63480 & 5.68838 & 0.00054 & -5.04930 & 0.00450 \\ 
        55506.59700 & 5.68795 & 0.00051 & -5.04760 & 0.00410 \\ 
        55512.60490 & 5.68950 & 0.00050 & -5.06380 & 0.00420 \\ 
        55516.62579 & 5.68935 & 0.00046 & -5.06690 & 0.00370 \\ 
        55524.58469 & 5.68953 & 0.00058 & -5.04520 & 0.00440 \\ 
\end{longtable}
}

\onllongtab{5}{
\begin{longtable}{ccccc}
\caption{\label{test} RVs and activity index for HD7449.}\\ 
\hline
\hline
Julian date  & RV & RV error & log(R'$_{HK}$) & log(R'$_{HK}$) error\\ 
$[\mathrm{T} - 2400000]$ & $[\mathrm{km}\,\mathrm{s}^{-1}]$ & $[\mathrm{km}\,\mathrm{s}^{-1}]$ & - & -\\ 
\hline
\endfirsthead
\caption{Continued.} \\ 
\hline
Julian date  & RV & RV error & log(R'$_{HK}$) & log(R'$_{HK}$) error\\ 
$[\mathrm{T} - 2400000]$ & $[\mathrm{km}\,\mathrm{s}^{-1}]$ & $[\mathrm{km}\,\mathrm{s}^{-1}]$ & - & -\\ 
\hline
\endhead
\hline
\endfoot
\hline
\endlastfoot
        52945.66121 & -19.75759 & 0.00130 & -4.85880 & 0.00550 \\ 
        53217.92351 & -19.77253 & 0.00068 & -4.85190 & 0.00300 \\ 
        53262.80762 & -19.77692 & 0.00053 & -4.86330 & 0.00200 \\ 
        53270.76277 & -19.76983 & 0.00054 & -4.82820 & 0.00200 \\ 
        53272.75520 & -19.76925 & 0.00066 & -4.84100 & 0.00280 \\ 
        53291.70634 & -19.77026 & 0.00082 & -4.84430 & 0.00380 \\ 
        53307.67324 & -19.77744 & 0.00066 & -4.86560 & 0.00300 \\ 
        53308.72059 & -19.77620 & 0.00054 & -4.85580 & 0.00220 \\ 
        53321.67138 & -19.77402 & 0.00057 & -4.83950 & 0.00200 \\ 
        53946.84580 & -19.76933 & 0.00070 & -4.85750 & 0.00270 \\ 
        53951.89653 & -19.76814 & 0.00100 & -4.86140 & 0.00410 \\ 
        54054.71440 & -19.73268 & 0.00055 & -4.83550 & 0.00180 \\ 
        54078.60172 & -19.74698 & 0.00066 & -4.87050 & 0.00240 \\ 
        54080.61805 & -19.74997 & 0.00063 & -4.88390 & 0.00230 \\ 
        54082.63518 & -19.74896 & 0.00060 & -4.86340 & 0.00210 \\ 
        54084.62233 & -19.73678 & 0.00080 & -4.85730 & 0.00330 \\ 
        54117.57988 & -19.75313 & 0.00070 & -4.84520 & 0.00320 \\ 
        54342.75820 & -19.76996 & 0.00061 & -4.85770 & 0.00200 \\ 
        54343.87281 & -19.77101 & 0.00069 & -4.86360 & 0.00250 \\ 
        54345.78963 & -19.77259 & 0.00053 & -4.85540 & 0.00150 \\ 
        54348.80066 & -19.76826 & 0.00056 & -4.83640 & 0.00220 \\ 
        54349.80995 & -19.76414 & 0.00058 & -4.82370 & 0.00170 \\ 
        54350.82673 & -19.76273 & 0.00062 & -4.82210 & 0.00200 \\ 
        54672.86349 & -19.76952 & 0.00058 & -4.83980 & 0.00180 \\ 
        54674.89475 & -19.76593 & 0.00083 & -4.85180 & 0.00320 \\ 
        54678.88163 & -19.76944 & 0.00093 & -4.83210 & 0.00360 \\ 
        54681.91007 & -19.77129 & 0.00082 & -4.84420 & 0.00300 \\ 
        54702.86374 & -19.76735 & 0.00067 & -4.84770 & 0.00230 \\ 
        54705.82511 & -19.76547 & 0.00079 & -4.83400 & 0.00280 \\ 
        54708.82073 & -19.76247 & 0.00070 & -4.83200 & 0.00230 \\ 
        54730.76492 & -19.77139 & 0.00069 & -4.85610 & 0.00250 \\ 
        54733.72836 & -19.76679 & 0.00054 & -4.84690 & 0.00170 \\ 
        54734.72516 & -19.76663 & 0.00086 & -4.84100 & 0.00320 \\ 
        54737.68372 & -19.76662 & 0.00081 & -4.84260 & 0.00300 \\ 
        54743.65465 & -19.76769 & 0.00057 & -4.85160 & 0.00190 \\ 
        54745.65226 & -19.76931 & 0.00060 & -4.84440 & 0.00200 \\ 
        54749.81955 & -19.77115 & 0.00065 & -4.85150 & 0.00240 \\ 
        54754.69528 & -19.77218 & 0.00062 & -4.86050 & 0.00200 \\ 
        54758.67215 & -19.76367 & 0.00060 & -4.83960 & 0.00210 \\ 
        54759.66411 & -19.76325 & 0.00058 & -4.82900 & 0.00190 \\ 
        54760.67606 & -19.76361 & 0.00052 & -4.82000 & 0.00160 \\ 
        54761.69393 & -19.75874 & 0.00049 & -4.81910 & 0.00150 \\ 
        54762.66829 & -19.76373 & 0.00051 & -4.83170 & 0.00160 \\ 
        54776.65127 & -19.76041 & 0.00064 & -4.82800 & 0.00210 \\ 
        54780.59676 & -19.77232 & 0.00066 & -4.84480 & 0.00220 \\ 
        55020.91992 & -19.76819 & 0.00050 & -4.86690 & 0.00160 \\ 
        55022.87206 & -19.76855 & 0.00066 & -4.87530 & 0.00250 \\ 
        55023.92477 & -19.76887 & 0.00060 & -4.87530 & 0.00210 \\ 
        55036.89860 & -19.76640 & 0.00065 & -4.87020 & 0.00230 \\ 
        55037.88268 & -19.77008 & 0.00052 & -4.87200 & 0.00170 \\ 
        55038.89187 & -19.77085 & 0.00054 & -4.87400 & 0.00180 \\ 
        55064.84770 & -19.76453 & 0.00057 & -4.85500 & 0.00180 \\ 
        55066.85709 & -19.76532 & 0.00060 & -4.86260 & 0.00200 \\ 
        55070.78553 & -19.76047 & 0.00063 & -4.85470 & 0.00220 \\ 
        55073.77085 & -19.75893 & 0.00059 & -4.83430 & 0.00200 \\ 
        55075.75263 & -19.76073 & 0.00083 & -4.84140 & 0.00360 \\ 
        55077.78739 & -19.76292 & 0.00075 & -4.86180 & 0.00310 \\ 
        55097.77727 & -19.76448 & 0.00058 & -4.86540 & 0.00190 \\ 
        55100.72145 & -19.75986 & 0.00068 & -4.84700 & 0.00240 \\ 
        55106.66702 & -19.75928 & 0.00065 & -4.85660 & 0.00220 \\ 
        55153.60187 & -19.75710 & 0.00075 & -4.86480 & 0.00260 \\ 
        55162.60845 & -19.75304 & 0.00065 & -4.83120 & 0.00210 \\ 
        55373.92311 & -19.71966 & 0.00094 & -4.83550 & 0.00490 \\ 
        55375.93913 & -19.72440 & 0.00079 & -4.86260 & 0.00360 \\ 
        55397.89627 & -19.73135 & 0.00109 & -4.86310 & 0.00520 \\ 
        55400.91531 & -19.72969 & 0.00061 & -4.85080 & 0.00230 \\ 
        55404.85328 & -19.73224 & 0.00068 & -4.84990 & 0.00240 \\ 
        55412.87128 & -19.73287 & 0.00077 & -4.86260 & 0.00320 \\ 
        55455.69289 & -19.73158 & 0.00072 & -4.85450 & 0.00280 \\ 
        55457.74651 & -19.73692 & 0.00069 & -4.84990 & 0.00260 \\ 
        55463.85029 & -19.72860 & 0.00072 & -4.83790 & 0.00290 \\ 
        55465.82382 & -19.72569 & 0.00080 & -4.83630 & 0.00320 \\ 
        55480.70492 & -19.72722 & 0.00061 & -4.83080 & 0.00210 \\ 
        55484.67754 & -19.73462 & 0.00057 & -4.84300 & 0.00200 \\ 
        55488.70552 & -19.73311 & 0.00058 & -4.84560 & 0.00200 \\ 
        55494.71331 & -19.72698 & 0.00070 & -4.85120 & 0.00260 \\ 
        55497.65285 & -19.73299 & 0.00066 & -4.84370 & 0.00230 \\ 
        55499.66954 & -19.73068 & 0.00087 & -4.84200 & 0.00340 \\ 
        55516.64347 & -19.73142 & 0.00072 & -4.84660 & 0.00270 \\ 
        55523.61990 & -19.72749 & 0.00072 & -4.85590 & 0.00280 \\ 
        55574.56674 & -19.72588 & 0.00058 & -4.84980 & 0.00220 \\ 
        55578.53207 & -19.72807 & 0.00067 & -4.83040 & 0.00230 \\ 
\end{longtable}
}

\onllongtab{6}{
\begin{longtable}{ccccc}
\caption{\label{test} RVs and activity index for HD137388.}\\ 
\hline
\hline
Julian date  & RV & RV error & log(R'$_{HK}$) & log(R'$_{HK}$) error\\ 
$[\mathrm{T} - 2400000]$ & $[\mathrm{km}\,\mathrm{s}^{-1}]$ & $[\mathrm{km}\,\mathrm{s}^{-1}]$ & - & -\\ 
\hline
\endfirsthead
\caption{Continued.} \\ 
\hline
Julian date  & RV & RV error & log(R'$_{HK}$) & log(R'$_{HK}$) error\\ 
$[\mathrm{T} - 2400000]$ & $[\mathrm{km}\,\mathrm{s}^{-1}]$ & $[\mathrm{km}\,\mathrm{s}^{-1}]$ & - & -\\ 
\hline
\endhead
\hline
\endfoot
\hline
\endlastfoot
        53577.53993 & 26.34614 & 0.00082 & -4.89960 & 0.00720 \\ 
        53814.83670 & 26.34973 & 0.00051 & -4.91800 & 0.00320 \\ 
        54200.85902 & 26.34939 & 0.00078 & -4.96370 & 0.00720 \\ 
        54202.82649 & 26.34877 & 0.00059 & -4.94330 & 0.00420 \\ 
        54562.83437 & 26.34819 & 0.00059 & -4.96050 & 0.00440 \\ 
        54563.80879 & 26.34682 & 0.00063 & -4.96960 & 0.00520 \\ 
        54564.80258 & 26.34689 & 0.00088 & -4.93990 & 0.00800 \\ 
        54566.81858 & 26.34410 & 0.00069 & -4.94740 & 0.00550 \\ 
        54570.77218 & 26.34473 & 0.00073 & -4.95220 & 0.00610 \\ 
        54571.79006 & 26.34402 & 0.00101 & -4.93210 & 0.01020 \\ 
        54610.83689 & 26.34066 & 0.00094 & -4.98850 & 0.01060 \\ 
        54611.66033 & 26.33828 & 0.00073 & -4.96380 & 0.00650 \\ 
        54612.72229 & 26.34350 & 0.00111 & -4.97030 & 0.01270 \\ 
        54616.65465 & 26.33977 & 0.00087 & -4.96480 & 0.00880 \\ 
        54617.74429 & 26.33999 & 0.00095 & -4.92290 & 0.01000 \\ 
        54638.78441 & 26.34646 & 0.00108 & -5.02560 & 0.01500 \\ 
        54639.71165 & 26.34637 & 0.00071 & -4.96330 & 0.00620 \\ 
        54640.67818 & 26.34629 & 0.00098 & -4.98620 & 0.01140 \\ 
        54641.67279 & 26.34560 & 0.00091 & -4.97670 & 0.00930 \\ 
        54642.64940 & 26.34429 & 0.00073 & -4.97330 & 0.00680 \\ 
        54643.66951 & 26.34162 & 0.00093 & -4.98000 & 0.01000 \\ 
        54644.67698 & 26.34264 & 0.00077 & -4.96760 & 0.00730 \\ 
        54646.58144 & 26.34221 & 0.00079 & -4.96870 & 0.00740 \\ 
        54649.45408 & 26.34157 & 0.00084 & -4.97520 & 0.00820 \\ 
        54679.55997 & 26.34599 & 0.00120 & -5.02140 & 0.01680 \\ 
        54950.81289 & 26.34607 & 0.00069 & -4.87880 & 0.00490 \\ 
        54952.79030 & 26.34498 & 0.00083 & -4.89000 & 0.00590 \\ 
        54953.76599 & 26.34792 & 0.00077 & -4.85890 & 0.00470 \\ 
        54954.77898 & 26.34987 & 0.00063 & -4.86360 & 0.00300 \\ 
        54988.65962 & 26.35016 & 0.00068 & -4.89050 & 0.00470 \\ 
        54991.69507 & 26.35100 & 0.00083 & -4.88330 & 0.00640 \\ 
        55025.53574 & 26.35193 & 0.00131 & -4.91810 & 0.01480 \\ 
        55038.50812 & 26.35479 & 0.00081 & -4.88540 & 0.00600 \\ 
        55039.58410 & 26.35372 & 0.00065 & -4.87200 & 0.00440 \\ 
        55224.88455 & 26.35618 & 0.00064 & -4.88070 & 0.00410 \\ 
        55227.87844 & 26.35394 & 0.00064 & -4.87550 & 0.00400 \\ 
        55241.88975 & 26.34778 & 0.00057 & -4.88180 & 0.00340 \\ 
        55242.87954 & 26.34918 & 0.00072 & -4.88760 & 0.00510 \\ 
        55243.88539 & 26.35039 & 0.00086 & -4.90390 & 0.00700 \\ 
        55246.86034 & 26.34978 & 0.00080 & -4.86900 & 0.00580 \\ 
        55247.87606 & 26.35105 & 0.00080 & -4.89740 & 0.00630 \\ 
        55280.84147 & 26.34495 & 0.00067 & -4.90030 & 0.00480 \\ 
        55283.80421 & 26.34576 & 0.00070 & -4.88930 & 0.00490 \\ 
        55370.53836 & 26.35431 & 0.00102 & -4.91720 & 0.01110 \\ 
        55372.59604 & 26.35266 & 0.00073 & -4.87830 & 0.00600 \\ 
        55374.56711 & 26.35676 & 0.00120 & -4.87780 & 0.01300 \\ 
        55403.52937 & 26.35504 & 0.00132 & -4.88790 & 0.01450 \\ 
        55412.53927 & 26.36110 & 0.00090 & -4.87860 & 0.00830 \\ 
        55424.49085 & 26.35933 & 0.00129 & -4.88700 & 0.01470 \\ 
        55437.50335 & 26.35928 & 0.00076 & -4.87930 & 0.00620 \\ 
        55438.48273 & 26.35856 & 0.00066 & -4.87020 & 0.00520 \\ 
        55453.47667 & 26.36861 & 0.00065 & -4.82730 & 0.00470 \\ 
        55456.48157 & 26.36404 & 0.00086 & -4.83240 & 0.00700 \\ 
        55460.50686 & 26.36353 & 0.00151 & -4.92160 & 0.02230 \\ 
        55464.56222 & 26.36450 & 0.00098 & -4.85710 & 0.00920 \\ 
        55591.86432 & 26.35511 & 0.00080 & -4.80940 & 0.00570 \\ 
        55611.85308 & 26.35044 & 0.00081 & -4.81620 & 0.00600 \\ 
        55615.85468 & 26.34877 & 0.00065 & -4.82940 & 0.00450 \\ 
        55620.89662 & 26.35022 & 0.00095 & -4.86440 & 0.00910 \\ 
        55624.90060 & 26.35480 & 0.00076 & -4.84110 & 0.00610 \\ 
        55628.89056 & 26.35172 & 0.00091 & -4.83630 & 0.00790 \\ 
        55631.88034 & 26.34628 & 0.00086 & -4.83900 & 0.00720 \\ 
\end{longtable}
}

\onllongtab{7}{
\begin{longtable}{ccccc}
\caption{\label{test} RVs and activity index for HD204941.}\\ 
\hline
\hline
Julian date  & RV & RV error & log(R'$_{HK}$) & log(R'$_{HK}$) error\\ 
$[\mathrm{T} - 2400000]$ & $[\mathrm{km}\,\mathrm{s}^{-1}]$ & $[\mathrm{km}\,\mathrm{s}^{-1}]$ & - & -\\ 
\hline
\endfirsthead
\caption{Continued.} \\ 
\hline
Julian date  & RV & RV error & log(R'$_{HK}$) & log(R'$_{HK}$) error\\ 
$[\mathrm{T} - 2400000]$ & $[\mathrm{km}\,\mathrm{s}^{-1}]$ & $[\mathrm{km}\,\mathrm{s}^{-1}]$ & - & -\\ 
\hline
\endhead
\hline
\endfoot
\hline
\endlastfoot
        53341.53507 & 32.64114 & 0.00054 & -4.95710 & 0.00400 \\ 
        53344.53090 & 32.64180 & 0.00053 & -4.97430 & 0.00360 \\ 
        54054.56108 & 32.63307 & 0.00054 & -4.91780 & 0.00320 \\ 
        54228.87836 & 32.63327 & 0.00064 & -4.92420 & 0.00420 \\ 
        54292.85826 & 32.63618 & 0.00068 & -4.89960 & 0.00340 \\ 
        54296.76897 & 32.63552 & 0.00066 & -4.90680 & 0.00310 \\ 
        54389.56546 & 32.63918 & 0.00093 & -4.90380 & 0.00670 \\ 
        54390.56226 & 32.63859 & 0.00061 & -4.89480 & 0.00330 \\ 
        54732.66878 & 32.64276 & 0.00065 & -4.89570 & 0.00370 \\ 
        54738.66375 & 32.64424 & 0.00099 & -4.89610 & 0.00780 \\ 
        54746.57986 & 32.64464 & 0.00050 & -4.91070 & 0.00250 \\ 
        54749.57587 & 32.64387 & 0.00057 & -4.90580 & 0.00310 \\ 
        54774.55471 & 32.64187 & 0.00060 & -4.92270 & 0.00340 \\ 
        54933.91111 & 32.64240 & 0.00061 & -4.93040 & 0.00390 \\ 
        54941.91026 & 32.63909 & 0.00047 & -4.92840 & 0.00230 \\ 
        55001.90189 & 32.64072 & 0.00066 & -4.95010 & 0.00440 \\ 
        55020.84267 & 32.64187 & 0.00045 & -4.91790 & 0.00210 \\ 
        55048.79335 & 32.64173 & 0.00094 & -4.92970 & 0.00750 \\ 
        55066.74188 & 32.63995 & 0.00066 & -4.95530 & 0.00430 \\ 
        55095.64678 & 32.63971 & 0.00063 & -4.95040 & 0.00300 \\ 
        55101.61504 & 32.64120 & 0.00091 & -4.96780 & 0.00800 \\ 
        55152.55912 & 32.63804 & 0.00058 & -4.94430 & 0.00350 \\ 
        55162.52708 & 32.63948 & 0.00064 & -4.94770 & 0.00420 \\ 
        55373.81738 & 32.63519 & 0.00076 & -4.97660 & 0.00660 \\ 
        55375.78771 & 32.63182 & 0.00079 & -4.99230 & 0.00750 \\ 
        55397.84505 & 32.63174 & 0.00115 & -5.02890 & 0.01480 \\ 
        55408.85463 & 32.63335 & 0.00068 & -4.96270 & 0.00520 \\ 
        55409.83046 & 32.63498 & 0.00063 & -4.96360 & 0.00520 \\ 
        55450.65245 & 32.63233 & 0.00075 & -4.97460 & 0.00530 \\ 
        55453.64105 & 32.63315 & 0.00058 & -4.97400 & 0.00310 \\ 
        55456.57248 & 32.63276 & 0.00066 & -4.96550 & 0.00500 \\ 
        55487.60470 & 32.63388 & 0.00068 & -4.95880 & 0.00500 \\ 
        55489.53648 & 32.63285 & 0.00046 & -4.96460 & 0.00260 \\ 
        55494.57706 & 32.63534 & 0.00059 & -4.96450 & 0.00390 \\ 
        55521.53647 & 32.63508 & 0.00076 & -4.97790 & 0.00600 \\ 
\end{longtable}
}

\bibliographystyle{aa}
\bibliography{dumusque_bibliography}

\end{document}